\documentclass[sigconf,letterpaper]{acmart}
\usepackage{amsmath,amssymb,amsfonts}
\usepackage{graphicx}
\usepackage{textcomp}
\usepackage{float,array}
\usepackage{mwe,tikz}
\usepackage{amsmath}

\usepackage{amsthm}
\usepackage{float}
\usepackage{enumitem}
\usepackage{hyperref}

\usepackage{rotating}
\usepackage{amssymb}
\usepackage{algpseudocode}
\usepackage{algorithm}
\usepackage{epstopdf}

\usepackage{multirow}
\usepackage{subfig}
\usepackage{dblfloatfix} 

\usepackage[textsize=tiny]{todonotes}
\usepackage{bbm}
\usepackage{xspace}
\usepackage{amsmath}
\usepackage{graphicx}
\usepackage{amssymb}
\usepackage{mathtools}
\usepackage[compact]{titlesec}
\usepackage[margin=5pt,font=small,labelfont={rm,bf}]{caption}
\makeatletter
\renewcommand*\env@matrix[1][*\c@MaxMatrixCols c]{%
  \hskip -\arraycolsep
  \let\@ifnextchar\new@ifnextchar
  \array{#1}}
\makeatother

\newcommand\paraspace{\vspace*{0.5ex}}
\newcommand\parab[1]{\paraspace\noindent\textbf{#1}}
\newcommand\parae[1]{\paraspace\textbf{\textit{#1}}}

\newcommand{\ie}{\emph{i.e.,}\xspace}
\newcommand{\eg}{\emph{e.g.,}\xspace}

\newcommand{\secref}[1]{\S\ref{#1}}
\newcommand{\figref}[1]{Figure~\ref{#1}}
\newcommand{\tabref}[1]{Table~\ref{#1}}

\newcommand{\eqnref}[1]{Equation~\ref{#1}}

\makeatletter
\renewcommand{\verbatim@font}{\ttfamily\bfseries\footnotesize}
\makeatother

\def\BibTeX{{\rm B\kern-.05em{\sc i\kern-.025em b}\kern-.08emT\kern-.1667em\lower.7ex\hbox{E}\kern-.125emX}}

\makeatletter
\def\maxwidth{\ifdim\Gin@nat@width>\linewidth\linewidth\else\Gin@nat@width\fi}
\def\maxheight{\ifdim\Gin@nat@height>\textheight\textheight\else\Gin@nat@height\fi}
\makeatother
\setkeys{Gin}{width=\maxwidth,height=\maxheight,keepaspectratio}
\makeatletter
\g@addto@macro{\UrlBreaks}{\UrlOrds}
\makeatother

\makeatletter
\let\origsection\section
\let\origsubsection\subsection

\renewcommand\section{\@ifstar{\starsection}{\nostarsection}}
\renewcommand\subsection{\@ifstar{\starsubsection}{\nostarsubsection}}

\newcommand\sectionprelude{\vspace{0ex}}
\newcommand\sectionpostlude{\vspace{0ex}}
\newcommand\subsectionprelude{\vspace{0ex}}
\newcommand\subsectionpostlude{\vspace{0ex}}

\newcommand\nostarsection[1]{\sectionprelude\origsection{#1}\sectionpostlude}
\newcommand\starsection[1]{\sectionprelude\origsection*{#1}\sectionpostlude}

\newcommand\nostarsubsection[1]{\subsectionprelude\origsubsection{#1}\subsectionpostlude}
\newcommand\starsubsection[1]{\subsectionprelude\origsubsection*{#1}\subsectionpostlude}

\makeatother

\renewcommand\paraspace{\vspace*{0.5ex}}
\providecommand\parab[1]{\paraspace\noindent\textbf{#1}}
\providecommand\parae[1]{\paraspace\textbf{\textit{#1}}}

\apptocmd\normalsize{%
\abovedisplayskip=4pt
\abovedisplayshortskip=45pt
\belowdisplayskip=4pt
\belowdisplayshortskip=4pt
}{}{}

\renewcommand\footnotetextcopyrightpermission[1]{} % removes footnote with conference info
\setcopyright{none}
\begin{document}
\title{Rapid Top-Down Synthesis of Large-Scale IoT Networks}

\author{Pradipta Ghosh}
\affiliation{\institution{University of Southern California}}
\email{pradiptg@usc.edu}

\author{Jonathan Bunton}
\affiliation{\institution{University of California, Los Angeles}}
\email{jonathan.m.bunton@gmail.com}

\author{ Dimitrios Pylorof}
\affiliation{\institution{Georgetown University}}
\email{dpylorof@austin.utexas.edu}

\author{Marcos A. M. Vieira}
\affiliation{\institution{Universidade Federal de Minas Gerais}}
\email{mmvieira@dcc.ufmg.br}

\author{Kevin Chan}
\affiliation{\institution{US Army Research Laboratory}}
\email{kevin.s.chan.civ@mail.mil}

\author{Ramesh Govindan}
\affiliation{\institution{University of Southern California}}
\email{ramesh@usc.edu}

\author{Gaurav S. Sukhatme }
\affiliation{\institution{University of Southern California}}
\email{gaurav@usc.edu}

\author{Paulo Tabuada }
\affiliation{\institution{University of California, Los Angeles}}
\email{tabuada@ucla.edu}

\author{Gunjan Verma}
\affiliation{\institution{US Army Research Laboratory}}
\email{gunjan.verma.civ@mail.mil}

\renewcommand{\shortauthors}{P. Ghosh et. al.}

\begin{abstract}

  Advances in optimization and constraint satisfaction techniques, together with the availability of elastic computing resources,  have spurred interest in large-scale network verification and synthesis. Motivated by this, we consider the top-down synthesis of ad-hoc IoT networks for disaster response and search and rescue operations. This synthesis problem must satisfy complex and competing constraints: sensor coverage, line-of-sight visibility, and network connectivity. The central challenge in our synthesis problem is quickly \textit{scaling} to large regions while producing cost-effective solutions. We explore two qualitatively different representations of the synthesis problems satisfiability modulo convex optimization (SMC), and mixed-integer linear programming (MILP). The former is more expressive, for our problem, than the latter, but is less well-suited for solving optimization problems like ours. We show how to express our network synthesis in these frameworks, and, to scale to problem sizes beyond what these frameworks are capable of, develop a \emph{hierarchical synthesis} technique that independently synthesizes networks in sub-regions of the deployment area, then combines these. We find that, while MILP outperforms SMC in some settings for smaller problem sizes, the fact that SMC's expressivity matches our problem ensures that it uniformly generates better quality solutions at larger problem sizes.

\end{abstract}

    % We take the first step towards developing such network synthesis tools.  We propose two approaches for network synthesis that are based on contemporary optimization paradigms such as Satisfiability Modulo Convex programming, used to solve a formulation of the problem with mixed continuous and Boolean variables subject to both convex and logical constraints, and pure SAT programming fused with graph-theoretic techniques, used to solve a purely discrete formulation of the same problem.
    % For illustration, we consider a problem of placing sensors comprising a network with prescribed degrees coverage and connectivity. Given a map of the configuration space, which can include obstacles, the sensing and communication radii, and the desired coverage and connectivity degrees, the proposed methods can be used to synthesize a topology that satisfies all relevant constraints.  Using both a wide range of synthetic and realistic deployment scenario, we illustrate the efficacy of the proposed solutions and compare and contrast the two approaches. 
    % The proposed techniques are expected to be of use in top-down network design applications, at various scales, that involve relatively scarce resources, as opposed to density control and sensor activation schemes often encountered in the literature.

\thanks{Research reported in this paper was sponsored in part by the Army Research Laboratory under Cooperative Agreement W911NF-17-2-0196. The views and conclusions contained in this document are those of the authors and should not be interpreted as representing the official policies, either expressed or implied, of the Army Research Laboratory or the U.S. Government. The U.S. Government is authorized to reproduce and distribute reprints for Government purposes notwithstanding any copyright notation here on.}
\maketitle

\thispagestyle{plain}
\pagestyle{plain}

\section{Introduction}
\label{sec:intro}

Over the last few years, optimization and constraint satisfaction technologies have improved to the point where they can be applied to large-scale verification and synthesis tasks. This, coupled with the advent of cloud computing, has spurred recent work in verifying network configurations of campuses~\cite{Kazemian:2012:HSA,Batfish} and data centers~\cite{Zeng:2014:LDC}. Beyond verification, these technologies have been applied to the synthesis of network configurations as well~\cite{beckett2017network}. Ultimately, this line of work will result in the synthesis of correct-by-design networks.
 
\parab{The problem.}
In this paper, we extend this line of work by considering the \emph{synthesis of a specific class of network topologies, namely, ad-hoc IoT networks}. These networks, often deployed on-demand (in a \textit{deployment region}) during disaster response or search and rescue operations, must satisfy at least four \textit{correctness} constraints (\secref{sec:problem}). First, every point of the deployment region must be covered by a small number ($k$) of sensors (the \textit{coverage constraint}). This $k$-coverage permits trilateration (for $k\geq 3$) of objects of interest (vehicles, people) in the environment. Second, because sensors may have line-of-sight limitations, the synthesis must account for obstacles in the deployment region (\textit{visibility constraints}). Third, obstacles constrain the placement locations of sensors (\textit{placement constraint}). Lastly, the deployed nodes must form a connected network such that there exists a network path between every pair of nodes (we call this the \textit{connectivity constraint}). This is necessary to convey sensor information between nodes or from node to a base station (potentially via other intermediate nodes).

Beyond these correctness constraints, we impose \textit{performance} objectives. Motivated by our use case of rapidly-deployable IoT networks for disaster response or search and rescue, we require that the network synthesis  (for brevity, henceforth we will use ``network synthesis'' to refer to the specific synthesis problem described above) \textit{scale} to campus-area deployments while generating topologies on the order of \textit{tens of minutes}. (By contrast, planning and deployment timescales for longer-lived networks, such as datacenters, can be on the order of weeks or months, if not longer~\cite{Lifecycle}).  Finally, to reduce deployment cost, the resulting synthesized network must have an objective of \textit{using the fewest number of sensor nodes}.

While prior work (\secref{sec:related}) has explored the synthesis of wired networks~\cite{schlinker2015condor}, of sensor placement for sensing coverage~\cite{younis2008strategies,wang2007efficient}, to our knowledge, no prior work has explored joint synthesis of coverage and connectivity in obstructed environments at scale.

\parab{Goal and Approach.}
The general network synthesis problem is non-convex, since it needs to determine coverage of areas not occupied by obstacles, and these areas can be non-convex. Because non-convex programming is fundamentally hard to scale, and because, over the past decade, significant effort has focused on speeding up solvers for lower-complexity formulations such as satisfiability (SAT)~\cite{z3} and linear and quadratic mixed integer formulations~\cite{GUROBI}, in this paper we explore synthesis using these two qualitatively different technologies (\secref{s:netw-synth-appr}), and compare the quality of the solutions they produce. Specifically, we compare  mixed-integer linear programming (MILP) and Satisfiability Modulo Convex Optimization (SMC~\cite{shoukry2018smc}); the latter is a recently proposed technique that permits expression of convex constraints in addition to satisfiability constraints.

\parab{Challenges.}
MILP and SMC fall into different points in the performance vs. expressivity space. SMC can model convex constraints, so it is comparable in expressivity to mixed-integer convex programming~\cite{shoukry2018smc}, but MILP can only model integer linear constraints. In this sense, SMC is a closer fit for our problem. Our first challenge is to formulate the constraints described above in these two frameworks, without significantly impacting \textit{solution quality}. In our setting, we measure solution quality by \emph{coverage redundancy}, the ratio between the average number of sensors covering a point in the deployment region to $k$. Our second challenge is \textit{scaling}, by which we mean the ability to obtain solutions for larger problem sizes that either SMC or MILP, by themselves, can solve for with a given set of computing resources. In general, SMC, because it solves a \textit{decision problem} can scale better than MILP. However, our problem of minimizing the total number of sensors is more naturally expressed in an optimization framework like MILP. In that sense, MILP is a better fit for our scaling objective. The goal of our paper is to understand how this tension between expressivity and scaling impacts solution quality.

%% SMC solvers iterate between a SAT solver and convex optimizers, and, for this reason, scale less well than the MILP. 

%% With the presence of thousands of IoT devices, future applications in megacities such as smart disaster management, personalized fine-grain navigation, or law enforcement operations are likely to involve a careful on-demand selection of the heterogeneous sensors (e.g., cameras, acoustic sensors) and actuators to form a network that can fulfill the application goals. This calls for a rapid, on-demand, goal-driven, agile, optimal, scalable, rapid network synthesis across these existing devices. 
%% Further, such synthesis needs to occur before the application takes place and must account for physical constraints, device characteristics, device failures, and the presence of adversaries. 
%% For example, a search and rescue operation in a megacities will require each location of a region on a congested city environment to be observed by at least one camera. 
%% One can either deploy a set of cameras to fulfill the requirement or leverage already deployed cameras or a combination of both. Now the question remains, \emph{how can we optimally select a subset of the existing devices as well as place additional devices if required?} This relatively recent problem is known to the networking community as the network synthesis problem.

\parab{Contributions.} To this end, our paper makes three contributions.

Our first contribution is an encoding of constraints in SMC (\secref{s:smc-form-netw}). To address the first challenge discussed above, we employ a \textit{grid-based} discretization of space, in which a grid is either occupied by an obstacle or not. Then, the coverage problem reduces to determining coverage of unoccupied grids; SMC can easily express these constraints. It can capture visibility using geometric constraints expressed in terms of half-planes that bound occupied grids, and network connectivity using constraints on a multi-hop adjacency matrix. Thus, SMC is able to jointly solve for coverage, visibility, and connectivity.

Our second contribution is an MILP formulation (\secref{s:basel-milp-form}), which also employs discretization, but transforms the visibility and coverage constraints into a graph vertex cover problem. The solution to this vertex cover problem is the placement of sensors that satisfies the visibility and coverage constraints, but the resulting network may not be connected. Expressing the (convex) connectivity constraint is much harder in MILP. However, using the observation that the second eigenvalue of the Laplacian of the connectivity graph determines whether the network is connected or not, we develop an affine inequality relaxation of the Laplacian that can encode the connectivity constraint in a MILP. If the network is not connected, we propose a novel approach to \textit{repair} missing connectivity (alternate to the Laplacian connectivity approach) from the coverage solution instead of a joint formulation.

Our third contribution addresses the scaling challenge by developing a \textit{ hierarchical synthesis} technique, which partitions the deployment region in sub-areas, solves each sub-area independently, then, in a second step, adds additional \textit{relay} nodes, using SMC's connectivity formulation, to ensure the network connectivity constraint. Hierarchical synthesis permits two relaxations that can speed up the synthesis: (a) solving only for coverage in the sub-areas (since, often, coverage will result in a substantially connected network) and establishing connectivity in the second step, and (b) solving for $j<k$ coverage in each of the sub-areas (since sensors in one sub-area can potentially cover sensors in another area), then repairing coverage in the second step.

% \pradipta{Have to Emphasize on the fact that we are the first to use SMC apart from the original paper. THis is a new technique etc.}

% Our third contribution develops an MILP formulation (\secref{s:basel-milp-form}) for synthesis. This re-uses the coverage and visibility constraint framing from SAT, but develops a novel technique to ensure the connectivity constraint. Using the observation that the second eigenvalue of the Laplacian of the connectivity graph can be used to determine connectivity, we develop an affine inequality relaxation of the Laplacian that can be used to encode the connectivity constraint in a MILP.

\parab{Summary of Results.} The performance of these two approaches is sensitive to the \textit{extent} of obstacles (\ie the fraction of the deployment region covered by obstacles) and their \textit{dispersion} (how the obstacles are spread across the region). For small deployment regions in which MILP does not require hierarchical synthesis, but SMC does, there are a few obstacle settings in which MILP has lower coverage redundancy (\ie a better solution quality) than SMC. However, for larger deployment regions in which both SMC and MILP need hierarchical synthesis to scale, SMC's coverage  redundancy is almost always better than MILP. Thus, even though MILP scales better, because it is less expressive than SMC, its connectivity and coverage approximations hurt the quality of its solutions at larger problem sizes. Beyond these synthetic deployment regions, we also evaluate our approaches on realistic campus deployments of about 0.75~km by 0.75~km. For these SMC is able to find solutions with 50\% lower coverage redundancy than MILP in about two minutes.

\section{The Network Synthesis Problem}
\label{sec:problem}

In this section, we formulate the network synthesis problem and its associated challenges.

\parab{Problem.}
Assume that we have a \textit{deployment region} (the area in which to deploy the network) $\mathbf{L} \subset \mathbb{R}^2$. Let $\mathbf{O}  \subset \mathbf{L}$ be the set of obstacles in the deployment region.
Suppose we have $N$ omni-directional sensors (\eg 360-degree cameras, acoustic sensors), and each sensor with location $s_{i} \in  \mathbb{R}^2$ has a sensing radius $r^s_i$ and a communication radius $r^c_{i}$.

The network synthesis problem seeks to \emph{find locations} $\mathbf{S} =\{s_{i} \in  \mathbb{R}^2  | i \in \{1,\cdots, N\}\}$ for the $N$ sensors such that:
\begin{enumerate}
\item At least $k$ sensors cover each point $l\in \mathbf{L} \setminus \mathbf{O}$.
\item The sensors form a connected network: \ie there exists a path from each node to every other node in the network.
\item The network uses as few of the $N$ sensors as possible.
\end{enumerate}

%% With these assumption, the network synthesis problems is: \emph{Given a set of $N$ sensors and the region of interest $\mathbf{L}$, what is the placement the sensors so that every location, $l\in \mathbf{L}$ is covered by at-least $k$ sensors and the network is connected, if feasible? What is the smallest values of $N$ to satisfy the requirements?}

% To simplify the rest of the description, we set $k=1$; I
% In subsequent sections, we explain the more general version of the problem.

%% Let us take $k=1$ just for illustration. To formulate the problem we need the following pieces first.

To concretely represent this synthesis problem, we need to model constraints imposed by limited sensing and communication ranges, as well as by visibility constraints in the environment. The following sub-paragraphs do this. 
% \pradipta{Maybe add a sample image illustration of the objective.}

\parab{Coverage constraint (C1).}
Let $\mathrm{dist}(x_i, x_j) = \| x_i-x_j \|_2$ represent the Euclidean distance between  $x_i, x_j \in \mathbb{R}^2$, any two locations in the deployment region $\mathbf{L}$.  The predicate $\mathcal{C}(s_i, l_j): \mathbb{R}^2 \times \mathbb{R}^2 \rightarrow \{0,1\}$ represents the fact that sensor $s_i\in \mathbf{S}$ \emph{covers} location $l_j\in \mathbf{L} \setminus \mathbf{O}$: 
\begin{equation}
\label{e:coverage}
 \mathcal{C}(s_i, l_j) = \big( \mathrm{dist}(s_i,l_j) \leq r^s_i \big) \bigwedge \mathcal{B}(s_i, l_j)
\end{equation}
where $r^s_i$ is the coverage radius of the sensor $s_i$.  $ \mathcal{B}(s_i, l_j)$ is a predicate that checks whether there exist a line of sight between $s_i$ and $l_j$ (defined below). Line-of-sight visibility constraints are important for sensors such as cameras. Furthermore, in practice, cameras have finite range because of the finite resolution of the camera itself: beyond a certain distance, objects become too small to be distinguishable to the human eye~\cite{wang2004image}. 

The $k$-coverage goal then reduces to:
\begin{equation}
 \sum_{i=1}^N \mathcal{C}(s_i, l_j) \geq k  \ \ \ \forall \ \  l_j \in \mathbf{L} \setminus \mathbf{O}
\end{equation}
This ensures that at least $k$ sensors cover each location within the deployment region but not within an obstacle.

% A other two types of representation possible are convex representation and polygon representation. 
% Box representation is a subset of convex/polygon representation. 
% In general, the convex representation of obstacles would be possisimilar to the box estimation as illustrated in \figref{fig:convex}. Even convex representation might over-approximate the shape of the object which can be again compensated by sub-diving the obstacle in smaller convex objects. However, such convex representation has some challenges related the synthesis problem that we discuss later.

\parab{Environmental Visibility Constraints (C2).} Obstacles pose a significant challenge for network synthesis. We explore synthesis for planar surfaces for which it suffices to model obstacles in two dimensions. Future work can generalize this to 3-D models of the environment.

Let $\mathcal{B}^o(s_i, l_j, o)$ be a predicate that checks whether obstacle $o$ does \textit{not} block the line of sight between $s_i$ and $l_j$. Then, we can model the visibility between $s_i$ and $l_j$, defined by the predicate $\mathcal{B}(s_i, l_j)$ as:
\begin{equation}
\begin{split}
    \mathcal{B}(s_i, l_j) &= \bigwedge_{o \in \mathbf{O}} \mathcal{B}^o(s_i, l_j, o)
\end{split}
\label{eqn:visi}
\end{equation}

\parab{Network connectivity constraint (C3).} If predicate $\mathcal{P}(s_i, s_j): \mathbb{R}^2 \times \mathbb{R}^2 \rightarrow \{0,1\}$ represents the direct connectivity between sensors $s_i, s_j\in \mathbf{S}$, then:
% \begin{equation}
%     \mathcal{P}(s_i, s_j) = \begin{cases} 1 \implies dist(s_i,s_j) \leq r^c_{i,j},   \\ 0 \implies dist(s_i,s_j) > r^c_{i,j} \mbox{\ or\ } i = j.\end{cases}
% \end{equation}
\begin{equation}
    \mathcal{P}(s_i, s_j) = \Big( dist(s_i,s_j) \leq r^c_{i,j} \Big)
\end{equation}
where $r^c_{i,j} = \min \{r^c_{i}, r^c_{j}\}$ and $r^c_{i}$ is the communication radius of sensor $s_i$. In this, we make two simplifying assumptions: that line-of-sight is not required for RF communication, and that wireless propagation follows a disk model. We have left to future work to relax these assumptions, because they can significantly impact the scaling of network synthesis. With these assumptions, and with a conservative choice of the communication radius, as we show later, we are able to synthesize functional networks that, when deployed in practice, satisfy the constraints. 

The predicate $\mathcal{P}(s_i,s_j)$ alone is not sufficient to establish \textit{network connectivity}. To do this, let $\hat{\mathbf{A}}_{N-1}$ be a matrix in which the $(i,j)$-th element represents the number of (N-1) hop paths without loops between sensors $i$ and $j$. Then, network connectivity holds if there is at least one path between each pair of nodes:
\begin{equation}
    (\hat{\mathbf{A}}_{N-1})_{ij} > 0 \ \ \  \forall i,j\in \{1, \cdots, N\} \ \ \mathrm{with} \ \  i \neq j.
 \label{eq:conn}
  \end{equation}

\parab{Obstacles and sensor placement (C4).}
To prevent the synthesizer from placing sensor $s_i$ on obstacles, let $\mathcal{V}^o\big(s_i, o\big)$ be a predicate that evaluates to true if $s_i$ is \textit{not} placed on obstacle $o$. Then, if $\mathcal{V}(s_i)$ is a predicate that checks whether sensor $s_i \in \mathbf{S}$ is not placed on any obstacle:
\begin{equation}
    \mathcal{V}\big(s_i\big) = \bigwedge_{o \in \mathbf{O}} \mathcal{V}^o\big(s_i, o\big).
\label{eqn:placement}
\end{equation}

\parab{The overall formulation.} Given this formulation, our network synthesis formulation reduces to finding the smallest number of sensors $N$ that satisfy \textit{all} four of the constraints listed in \tabref{tab:prob_overall}.

{\small
\begin{table}[t]
    \centering
     \caption{Summary of Problem Formulation}
    {
    \begin{tabular}{ccc}
            \hline \vspace{1ex}
        Coverage & (C1)  &  $\sum_{i=1}^N \mathcal{C}(s_i, l_j) \geq k  \ \ \ \forall \ \  l_j \in \mathbf{L} \setminus \mathbf{O} $ \\
           
        Visibility & (C2)  &  $\mathcal{B}(s_i, l_j) = \bigwedge_{o \in \mathbf{O}} \mathcal{B}(s_i, l_j, o)$\\\vspace{1ex}
            & &$\forall s_i \in \mathbf{S}$  and $l_j\in \mathbf{L} \setminus \mathbf{O}$\\ 
            
        Connectivity & (C3)  &  $(\hat{\mathbf{A}}_{N-1})_{ij} > 0 $\\ \vspace{1ex}
        & & $\ \ \  \forall i,j\in \{1, \cdots, N\} \ \ and \ \  i \neq j$\\ 
        Placement & (C4)  & $ \mathcal{V}\big(s_i\big) \ \ \  \forall s_i \in \mathbf{S}$\\ \hline
    \end{tabular}}
    \label{tab:prob_overall}
\end{table}
}
\section{Approaches to Network Synthesis}
\label{s:netw-synth-appr}

\parab{Performance Goals.} \tabref{tab:prob_overall} lists constraints on the \textit{correctness} of network synthesis. In addition, motivated by the problem of quick, ad-hoc deployments of IoT networks (\secref{sec:intro}) in large deployment areas, we impose two performance objectives: synthesizing a network, within \textit{tens of minutes}, to cover a large urban campus of a \textit{1-2 sq. kms.} As we show in the rest of the paper, these \textit{scaling goals} stress the capabilities of existing synthesis methods.

\parab{Synthesis approaches.} Recent advances in fast constraint satisfaction and optimization permit the use of formal methods for network synthesis, and the goal of this paper is to understand how well these approaches can satisfy our correctness constraints and performance goals. Specifically, in this paper, we compare two qualitatively different approaches to network synthesis: one built upon a \textit{constraint satisfaction} framework called SMC~\cite{shoukry2018smc}, and another on a traditional \textit{optimization} approach.

\parab{Satisfiability Modulo Convex (SMC) Theory.} SMC extends Boolean Satisfiability (SAT). A Boolean Satisfiability (SAT) problem finds feasible assignments to Boolean variables consistent with a set of constraints, typically represented as the conjunction of a set of Boolean clauses, for example:
\begin{equation}
    a_1 \wedge (a_2 \vee a_3) \wedge (a_1 \vee  a_3)
    \label{eqn:satexample}
\end{equation}
Here, $a_1, a_2, a_3$ are Boolean variables. A SAT solver attempts to find an assignment for the Boolean variables, such that the entire clause evaluates to \textsc{True} given the formula is satisfiable. For example, \eqnref{eqn:satexample} can be satisfied the following assignment: $a_1 = \textsc{True} , a_2 =\textsc{False}, a_3 = \textsc{True}$, and the SAT solver returns this assignment. On the other hand, if no such assignment exists, the function expressed by the formula is \textsc{False} for all possible variable assignments and the formula is unsatisfiable and the SAT solver identifies that no satisfying assignment exists. For example, $a_1 \wedge \neg a_1$ is unsatisfiable. SAT problems are typically hard to solve, but recent SAT solvers (\cite{z3,minisat}) scale well to problems of practical interest~\cite{fraisse2016boolean,lynce2006sudoku}.

% One can also represent integer programming and pseudo-Boolean constraints in terms of a SAT formulation~\cite{Biere:2009:HSV:1550723}, via straightforward transformations.

%\ramesh{Add a citation for SAT somewhere.}
% Marcos: added \cite{Biere:2009:HSV:1550723}

% For network synthesis, however, SAT cannot easily represent the network connectivity constraint (C3): in \secref{sec:proposed-sat}, we discuss how we address this challenge.

%% on fast Boolean Satisfiability problem (SAT), Satisfiability Modulo Convex (SMC), Mixed Integer Linear Programming (MILP) solvers has enabled us to move away from heuristics and adapt a more systematic and optimized approach. The core behind SAT approaches lie in representing a problem in terms of a set of Boolean Satisfiability clauses and searching for a solution to fulfill the requirements.

%% With recently developed solvers like z3, we can solve for very large set of Boolean constraints within couple of seconds to minutes. However, this requires discretization and approximations of all the constraints

SMC~\cite{shoukry2018smc}, designed to address the feasibility of mixed-integer convex problems, uses a SAT solver to suggest admissible assignments for a problem's Boolean variables\footnote{Note that one can almost trivially encompass integer variables via a larger number of Boolean variables, following the appropriate transformations.} and a convex solver (\eg \cite{CPLEX}) to suggest admissible values of the problem's real variables. To bridge the two tools, it represents convex constraints using \textit{pseudo-Boolean} variables.

Consider the following example:
\begin{equation}
\begin{split}
\end{split}
    a_1 \wedge (a_2 \vee a_3) \wedge (a_1 \rightarrow x_1 + x_2 = 4)
    \label{eqn:satexample2}
\end{equation}
Here, $a_1, a_2, a_3$ are Boolean variables and $x_1$,  $x_2$ are real variables. The third clause implies that if $a_1$ is \textsc{True}, we should be able to find a valid value of $x_1$ and  $x_2$ such that $x_1 + x_2 = 4$ (a convex constraint).

SMC replaces the convex constraint with a pseudo-Boolean variable, say $a_4$, then runs the SAT solver. If the solver produces a \textsc{True} assignment for both $a_4$ and $a_1$, then it attempts to use the convex solver to find a satisfying assignment for the convex constraint. If none are found (in our example, there exists a satisfying assignment), the output of the convex solver is used to produce a \textit{counter-example} to constrain the search space for the SAT solver. This leads to a search over a more constrained SAT solution space by excluding conflicting combinations of pseudo-boolean variables corresponding to the counter-example. This process repeats until the suggested combination of convex constraints is satisfiable.

When used for network synthesis, SMC produces a \textit{feasible} solution, given $N$, the number of sensors, as input (or indicates infeasibility). To synthesize a network with the fewest sensors, we perform a binary search on $N$.

\parab{Mixed-Integer Linear Programming (MILP).}
Optimization techniques like Mixed Integer Linear Programming (MILP) can  directly synthesize a network with the fewest number of sensors. MILP
is a version of integer programming in which variables need not all be integers, but the constraints need to be linear. The connectivity constraint (C3) is hard to express in MILP, and \secref{s:basel-milp-form} describes how we address this challenge.

%% Even so, as we discuss later, our MILP formulation scales worse than SAT and SMC based methods.

%% \tabref{tab:solvercomp} summarizes the qualitative distinctions between these approaches. In the rest of this paper, we explore quantitatively how well (and under what conditions) each approach scales.

%% In summary, all three methods can be used for representing the synthesis problem with some advantages and disadvantages, summarized in \tabref{tab:solvercomp}.
%% Thus, we decided to mainly use SAT and SMC method with a focus on improving the scalability. 
%% We have also developed a MILP version of network synthesizer but do not compare its performance as it stops scaling  even with a very small problem. 

%% \begin{table}[!ht]
%%     \centering
%%     \caption{Solver Technique Comparison}

%%     \begin{tabular}{c|c|c}
%%                         &SMC & MILP \\\hline
%%       Scalability Range  & Low  &    High \\
%%       Runtime            & High  &    Low \\
%%       Approximation     & Low  &    High \\

%%     \end{tabular}
%%     \label{tab:solvercomp}
%% \end{table}

%% \input{proposed-sat.tex}
%% \input{milp_pradipta.tex}
\section{SMC Formulation of Network Synthesis}
\label{s:smc-form-netw}

%% In \secref{sec:proposed-sat}, we use a highly restricted discrete approximation of the  network synthesis problem to encode into a SAT problem. In this section, we present our second scalable approach for network synthesis where the amount of approximation is much less than the discrete SAT approach.

In this section, we describe how we cast network synthesis in the SMC framework~\cite{shoukry2018smc}.

\begin{figure}[t]
    \centering
    \includegraphics[width=0.5\linewidth]{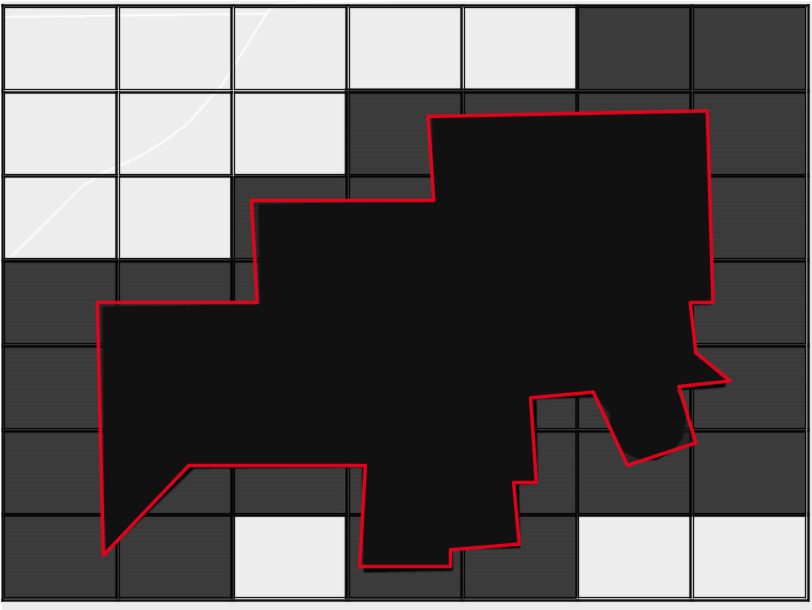}
    % \subfloat[]{\label{fig:visibilityrules2:1} \includegraphics[width=0.25\linewidth]{images/los_illius1.eps}} \quad
    % \subfloat[]{\label{fig:visibilityrules2:2}\includegraphics[width=0.25\linewidth]{images/los_illius2.eps}}
    %  \subfloat[]{\label{fig:convex}\includegraphics[width=0.48\linewidth]{images/convex_illus.png} 
    \caption{A complex obstacle can be modeled as a polygon, but this can be computationally difficult. We discretize space into grids and model each grid as occupied (if even a part of an obstacle overlaps with the grid) or unoccupied. Occupied grids are shaded in a dark color. }
    \label{fig:poly}
  \end{figure}

\parab{Area Coverage.}
The coverage constraint (C1, \eqnref{e:coverage}) specifies \textit{point} coverage, where the locations $l_j$ represent discrete locations in the deployment region $\mathbf{L}$. In practice, IoT deployments desire \textit{area} coverage, where at least $k$ sensors cover every point within $\mathbf{L}$ that is not covered by an obstacle. The area coverage problem in presence of obstacles is not convex, so it is hard to represent it using SMC (or, for that matter, MILP).
% \pradipta{Not 100\% sure about this statement. I mean area coverage problem can be cast into convex problem. }

\parae{Discretizing the deployment region.}
To address this challenge, we discretize space into a uniform grid which induces
a new discretized space $\mathbf{D}$. Each grid element $d \in \mathbf{D}$ can be represented in 2-D space as the \textit{intersection of four half planes}: 
\begin{equation}
\begin{split}
    \mathcal{I}^d(\mathbf{x}) = \mathcal{I}^d(x,y) & = (x > x^d_{min}) \wedge (x < x^d_{max}) \\
    & \wedge (y > y^d_{min}) \wedge (y< y^d_{max})
\end{split}
\label{eqn:halfplane}
\end{equation}
The respective four physical corners are $\mathbf{L}^d = \lbrace \left(x^d_{min},  y^d_{min}\right)$, $\left(x^d_{max},  y^d_{min}\right)$, $\left(x^d_{max},  y^d_{max}\right)$, $\left(x^d_{min},  y^d_{max}\right) \rbrace$.

\parae{Representing obstacles.} In general, we can represent an obstacle as a fine-grain polygon (\figref{fig:poly}). However, the finer the representation of the obstacles, the less scalable the network synthesis process becomes. A single obstacle with $P$ vertices, creates $\mathcal{O}(M\cdot N \cdot P)$ constraints, where $M$ represents the set of points covered by sensors. So, we leverage the space discretization to improve synthesis scaling: we say that each grid element $d\in\mathbf{D}$ is \textit{occupied} if an obstacle covers even a part of the element, else the element is \textit{unoccupied} (\figref{fig:poly}). Furthermore, since every $d\in\mathbf{D}$ is either occupied by an obstacle or free, $\mathbf{D}$ is the union of two disjoint sets $\mathbf{O}$ and $\mathbf{U}$, corresponding to occupied (obstacles) and unoccupied spaces, respectively. 

\begin{table}[t]
    \centering
    \caption{Pseudo Boolean for SMC Formulation}

    \resizebox{\linewidth}{!}{
    \begin{tabular}{cccccc}\hline
                    &   &   All &   Pseudo   & & Respective   \\
        Description &   &  Possible   &  Boolean  & &  Predicate or \\
                    &     &  Combinations &  Variable & & Constraints \\
            \hline
        Location   & $\psi_1=$& $ \bigwedge_{i=1}^{N} \bigwedge_{l_j \in \mathbf{L}}$ & $\Big(b^s_{ij}$  & $\rightarrow  $ &  $\mathcal{C}(s_i, l_j) \Big)$ \\     
        Coverage   &   & & & & \\\hline
        Grid   & $\psi_2 =$& $ \bigwedge_{u_g\in \mathbf{U}}$ & $\Big(b^u_{ig}$  & $\rightarrow  $&  $\bigwedge_{l_j \in \mathbf{L}^u_g} \mathcal{C}(s_i, l_j)\Big)$ \\
        Coverage & & & & &\\\hline
        Visibility     & $\psi_3 = $& $\bigwedge_{i=1}^{N} \bigwedge_{l_j \in \mathbf{L}} \bigwedge_{o_g \in \mathbf{O}}$  & $\Big(b^o_{ijg}$   & $\rightarrow  $ &  $\mathcal{B}^o(s_i, l_j, o_g) \Big)$ \\\hline
        
        Link  & $\psi_4 = $& $\bigwedge_{i=1}^{N} \bigwedge_{j=1, j \neq i}^{N}$&$\Big(b^c_{ij}$   & $\rightarrow  $ &  $ \mathcal{P}(s_i, s_j) \Big)$\\ 
        Connectivity &  & & & & \\\hline
        Placement & $\psi_5 = $& $\bigwedge_{i=1}^{N}  \bigwedge_{o_g \in \mathbf{O}}$& $\Big(b^v_{ig}$   & $\rightarrow   $ & $\mathcal{V}^o\big(s_i,  o_g\big) \Big) $ \\ 
         \hline
    \end{tabular}
    }
    \label{tab:pseudo_bools}
\end{table}

\parae{Defining area coverage.} If a sensor $s_i$ covers the four corners of a grid element $u_g  \in \mathbf{U}$, then it covers all points within the element. Thus, we can represent the problem of covering the entire area $\mathbf{L}$ by the problem of covering the corners of unoccupied grids. Formally, let $\mathbf{L} = \bigcup_{u_i \in \mathbf{U}} \mathbf{L}^u_i$  where $\mathbf{L}^u_i$ denotes the four corners of the grid $u_i$. Then, our formulation uses the pseudo-Boolean variables in \tabref{tab:pseudo_bools} to directly represent the convex and complex constraints (\secref{sec:problem}).

We can encode the $k$-coverage problem using SMC as follows.

\parae{Coverage Constraints (C1).}
We ensure that at least $k$ sensors cover each grid region $u_g$ as follows:
\begin{equation}
     \left( \sum_{i=1}^{N} b^u_{ig} \geq k \right) 
     \label{eqn:k-coverage}
\end{equation}
where covering a grid implies that the same sensor covers all four corners of the grid $(b^u_{ig} \rightarrow   \bigwedge_{l_j\in \mathbf{L}^u_g} b^s_{ij})$
which expresses the constraint that if sensor $s_i$ covers a grid $u_g$  ($b^u_{ig} = 1\ \ or \ \ \textsc{True}$), then it covers all four corners of the grid $l_j\in \mathbf{L}^u_g$.

\begin{figure}[!ht]
    \centering
    \subfloat[]{\label{fig:visibilityrules2:1} \includegraphics[width=0.4\linewidth]{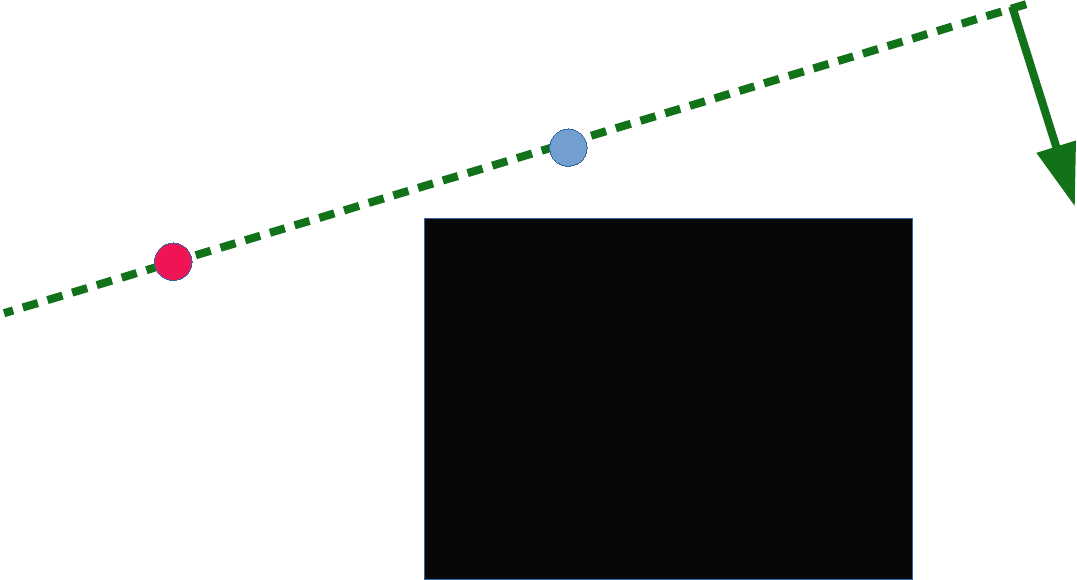}} \quad
    \subfloat[]{\label{fig:visibilityrules2:2}\includegraphics[width=0.4\linewidth]{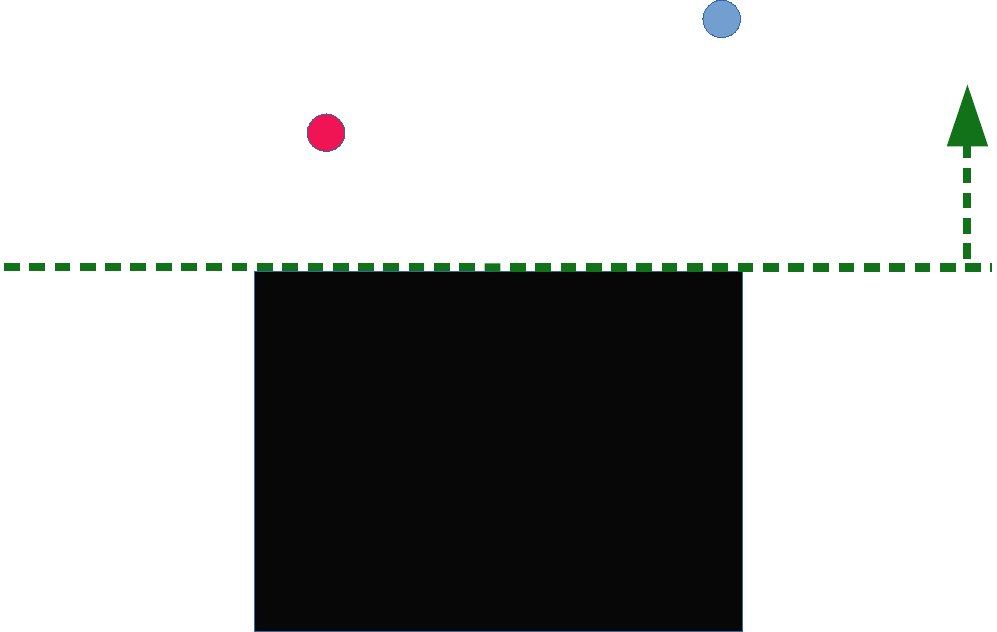}}
    \caption{Two visibility conditions. Red dot signifies a sensor and Blue dot signifies a sensed location.}
    \label{fig:visibilityrules2}
\end{figure}

\parae{Visibility Constraints (C2).} The four half-planes (\eqnref{eqn:halfplane}) of each occupied grid element can also model visibility. Let $\mathbf{L}^o_g$ represent the respective four corners of occupied grid element $o_g\in \mathbf{O}$. Then, line of sight depends upon two conditions: 
\begin{enumerate}
  \item[(v1)] This condition applies when all four vertices of an obstacle are on the same half plane created by the line joining a sensor location $s_i$ and a sensed location $l_j$ (\figref{fig:visibilityrules2:1}). If the line equation joining $s_i$, $l_j$ is $f_{ij}(\mathbf{x} =(x,y)) = 0$ then all four vertices should satisfy either $f_{ij}(\mathbf{x}) > 0$ or $f_{ij}(\mathbf{x}) < 0$, expressed as follows:
    {\small
    \begin{equation}
       \mathcal{B}_1(s_i, l_j, o_g) =  \Bigg( \bigwedge_{\mathbf{x}\in \mathbf{L}^o_g} (f_{ij}(\mathbf{x}) > 0)\Bigg) \bigvee \Bigg(\bigwedge_{\mathbf{x}\in \mathbf{L}^o_g} (f_{ij}(\mathbf{x}) < 0)\Bigg)
    \end{equation}
    }
    \item[(v2)] This condition applies when both the sensor location and the sensed location are on the outer half-plane of at least one obstacle face  (\figref{fig:visibilityrules2:2}). Mathematically, both points should satisfy one of the following four conditions: $x < x^o_{min}$, $x > x^o_{max}$, $y < y^o_{min}$, $y > y^o_{max}$, expressed as follows:
    
    {\small
     \begin{equation}
    \begin{split}
    &\mathcal{B}_2\Big(s_i = (x_1, y_1), l_j = (x_2, y_2), o_g\Big)  = \Big(x_1 < x^o_{min}\bigwedge x_2 < x^o_{min} \Big)\\
    & \bigvee  \Big(x_1 > x^o_{max} \bigwedge x_2 > x^o_{max} \Big)\bigvee \Big( y_1 < y^o_{min} \bigwedge y_2 < y^o_{min}\Big)  \\
    & \bigvee \Big( y_1 > y^o_{max} \bigwedge  y_2 > y^o_{max}\Big)
    \end{split}
    \label{eqn:visi2}
    \end{equation}
    }
    % \begin{equation}
    % \begin{split}
    % &\mathcal{B}_2\Big(s_i = (x_1, y_1), l_j = (x_2, y_2), o_g\Big)  = \\
    % &\Big((x_1 < x^o_{min})\wedge (x_2 < x^o_{min}) \Big) \bigvee \\ 
    % &\Big((x_1 > x^o_{max}) \wedge (x_2 > x^o_{max}) \Big)\bigvee   \\
    % &\Big( (y_1 < y^o_{min}) \wedge (y_2 < y^o_{min})\Big) \bigvee \\
    % &\Big( (y_1 > y^o_{max}) \wedge ( (y_2 > y^o_{max})\Big)
    % \end{split}
    % \label{eqn:visi2}
    % \end{equation}
  \end{enumerate}

% \ramesh{The superscript 2 above is not really a squaring, right, so use something else?}

% \begin{equation}
%     \begin{split}
%     &\mathcal{B}_2\Big(s_i = (x_1, y_1), l_j = (x_2, y_2), o_g\Big)  = \\
%     & \bigvee_{c\in \{ x^o_{min}, - x^o_{max}\}} ((x_1 < c) \wedge (x_2 < c))\\ 
%     &\bigvee_{c\in \{ y^o_{min}, - y^o_{max}\}} ((y_1 < c) \wedge (y_2 < c))
%     \end{split}
%     \label{eqn:visi2}
%     \end{equation}

If $\mathcal{B}^o(s_i, l_j, o_g)$ is a predicate that checks whether an obstacle grid element $o_g$ does not block the line of sight between $s_i$ and $l_j$, then we can write:
\begin{equation}
\begin{split}
    \mathcal{B}^o(s_i, l_j, o_g) &= \mathcal{B}_1(s_i, l_j, o_g) \vee \mathcal{B}_2(s_i, l_j, o_g) \\
\end{split}
\end{equation}
Using this, we can compute the visibility constraint (\eqnref{eqn:visi}).

% \\
%     \mathcal{B}_{O}(s_i, l_j) &= \bigwedge_{v \in o_g} \mathcal{B}_{F}(s_i, l_j, v) \\
% \ramesh{Do we need the subscript in $B_{O}$? It would be good to see if we can simplify the notation a little bit throughout the paper}.

%% To express line-of-sight visibility, we use convex constraints described in \secref{sec:problem}: $\mathcal{B}(s_i, l_j) = \bigwedge_{o_g \in \mathbf{O}} \mathcal{B}^o(s_i, l_j, o_g)$  which represents the condition that none of the obstacles $o_g \in \mathbf{O}$ obstructs the visibility between $s_i$ and $l_j$, resulting in:
%% \begin{equation}
%%   \mathcal{B}(s_i, l_j) = \bigwedge_{o_g \in \mathbf{O}} b^o_{ijg}
%% \end{equation}

\parae{Accommodating Heterogeneity.} If the network synthesis requires each location to be covered by multiple types of sensors such as cameras and microphones, we can incorporate that as follows: (1) to accommodate different sensing range coverage radii we can use different values of $r^s_i$ in \eqnref{e:coverage}, (2) to accommodate different visibility constraints, we can use a sensor-specific visibility function ($\mathcal{B}(s_i, l_j)$), and (3) to accommodate k-coverage constraints for different sensors, we can replicate \eqnref{eqn:k-coverage} for each type of sensors. This is possible because the SMC formulation treats each sensor as an independent element (even for the same type of sensors).

\parab{Connectivity Constraints (C3).} In a connected network, there should exist a communication path (single-hop or multihop) between every pair of nodes. Consider the adjacency matrix $\mathbf{A}$: $\mathbf{A}_{ij}  = b^c_{ij}$ where $b^c_{ij} \in \{0,1\}$ with True as 1 and False as 0. Now, let $\hat{A}_h$ represent the $h$-hop adjacency matrix (whose $i,j$-th entry is the number of $h$-hop paths between $s_i$ and $s_j$), then we can write the connectivity constraint as:
\begin{equation}
     \bigwedge_{j=2}^{N} \left( (\hat{\mathbf{A}}_{N-1})_{1j} > 0 \right).
   \end{equation}
This is a simplified version of (\eqnref{eq:conn}): we only check connectivity between node 1 and every other node because, if there exists a path from node 1 to every other node, they can always connect via node 1.

\parab{Obstacles and sensor placement (C4).}
To prevent the synthesizer from placing sensors in grid elements occupied by obstacles, we can define constraints that ensure that each sensor $s_i$'s position is outside the four half planes (\eqnref{eqn:halfplane}) defined by each grid element, as follows:
\begin{equation}
\begin{split}
    \mathcal{V}^o\big(s_i = (x,y), o_g\big) &= (x < x^o_{min}) \vee (x > x^o_{max}) \vee  \\
    & (y < y^o_{min}) \vee (y > y^o_{max})
\end{split}
% \label{eqn:visi}
\end{equation}
where $x^o_{min}, x^o_{max}, y^o_{min}, y^o_{max}$ define the four halfplanes of the obstacle grid. Finally, to make sure that the SMC solver does not put the sensor inside the obstacles, we add the following sets of clauses, the equivalent of (\eqnref{eqn:placement}):
\begin{equation}
     \bigwedge_{i=1}^{N}  \bigwedge_{o_g \in \mathbf{O}}  b^v_{ig} 
\end{equation}
This forces the solver to only select locations from the unoccupied grid elements.

\parab{The overall formulation.}
Putting all these together, the four key constraints can be summarized in \tabref{tab:prob_overall_smc}. From this, we arrive at the overall SMC formulation:  $\psi = \bigwedge_{i=1}^{8} \psi_i$. With this formulation, we perform a binary search on $N$ to find the solution with the fewest number of sensors.

{\small
\begin{table}[t]
    \centering
    \caption{Summary of SMC Encoding of k-Coverage}
    \resizebox{0.8\linewidth}{!}{
    \begin{tabular}{ccc}
            \hline
        Coverage  & (C1)  &  $\psi_6 =  \bigwedge_{u_g\in \mathbf{U}} \left( \sum_{i=1}^{N} b^u_{ig} \geq k \right)$\\     
        Visibility & (C2)  &  $ \bigwedge_{o_g \in \mathbf{O}} b^o_{ijg}  $ \\
        Connectivity & (C3)  & $\psi_7 = \bigwedge_{j=2}^{N} \left( (\hat{\mathbf{A}}_{N-1})_{1j} > 0 \right)$\\ 
        Placement & (C4)  &  $\psi_8 = \bigwedge_{i=1}^{N}  \bigwedge_{o_g \in \mathbf{O}}  b^v_{ig}$\\ \hline
    \end{tabular}}
    % \caption{Caption}
    \label{tab:prob_overall_smc}
  \end{table}
}

\parab{Constraint pre-computation.} Our SMC encoding discretizes the deployment region. If two such locations $l_i$ and $l_j$ are more than twice the sensing radius apart, then no single sensor can cover them. While SMC can eventually determine this through counter-examples with our encoding, we have found that it significantly improves the speed of synthesis to provide these as pre-computed constraints. We do this by (a) finding all pairs $l_i$ and $l_j$ that are greater than twice the sensing radius apart and (b) adding a constraint that prevents a single sensor from covering them. (We add similar constraints for connectivity as well).

\section{MILP-Based Network Synthesis}
\label{s:basel-milp-form}

In this section, we describe a scalable network synthesis method using a Mixed Integer Linear Programming (MILP) formulation. To use MILP for network synthesis, we need to find a \textit{MILP-encoding} of the constraints in \secref{sec:problem}.

%% To this end, we use the discretized space and for simplicity we assume that all nodes have same sensing radius $r^s$ and communication radius $r^c$.

%% The main advantage of a SAT formulation is that it permits adding boolean constraints to the formulation easily. \ramesh{It would be  good to add an example of an interesting constraint that could be expressed using boolean variables.}

%Next, we detail our SAT encoding of the k-coverage topology synthesis problem.
% It is worth mentioning that similarly to SMC (TODO: confirm that this appears after SMC section), SAT programming is dealing with constraint solving exclusively, that is, there is formally no objective function to be minimized or maximized, but only a logical combination of constraints to be satisfied. One can still use SAT in problem-specific optimization contexts (e.g. by trying to solve a problem with a variable corresponding to the objective function provided as a parameter, calling SAT as an oracle, and iterating or bisecting until the problem is not satisfiable). In the class of problems we consider in this work

\parab{Representing Coverage.}
To represent the coverage constraint C1 (\eqnref{e:coverage}) in MILP, we leverage the grid discretization and relax the definition of (a) distance and (b) visibility.

\parae{Distance relaxation.}
To reason about the constraints from \secref{sec:problem} in grid space, we first need to modify the distance function between any two grids in the unoccupied space $\mathbf{U}$ as follows:
\begin{equation}
    \mathrm{dist}'(u_i, u_j) = \max_{x_i\in \mathbf{L}^u_{i}, x_j \in  \mathbf{L}^u_{j}} \mathrm{dist}(x_i, x_j) \ \ , \  u_i, u_j \in \mathbf{U}
\end{equation}
where $\mathbf{L}^u_i$ represent the four corners of grid element $u_i\in \mathbf{U}$. This evaluates to the maximum distance between the two discrete regions. The largest distance measurements in this particular discrete space are between opposite corners of the rectangles describing $u_i$ and $u_j$. 

\begin{figure*}[h]
    \centering
    \subfloat[]{\label{fig:visibilityrules:1}\includegraphics[width=0.23\linewidth]{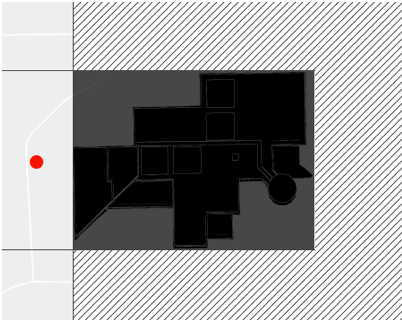}} \ 
    \subfloat[]{\label{fig:visibilityrules:2}\includegraphics[width=0.23\linewidth]{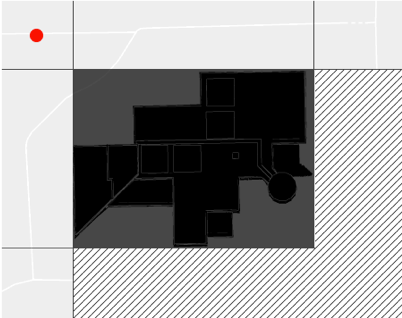}} \ 
    \subfloat[]{\label{vizgraph}\includegraphics[width=0.265\linewidth]{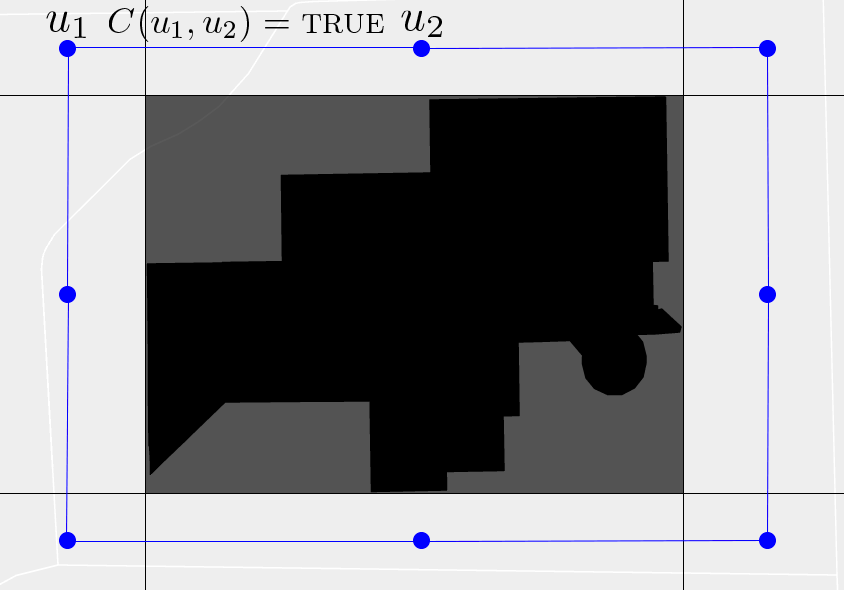}}\ 
    \subfloat[]{\label{collapse}\includegraphics[width=0.23\linewidth]{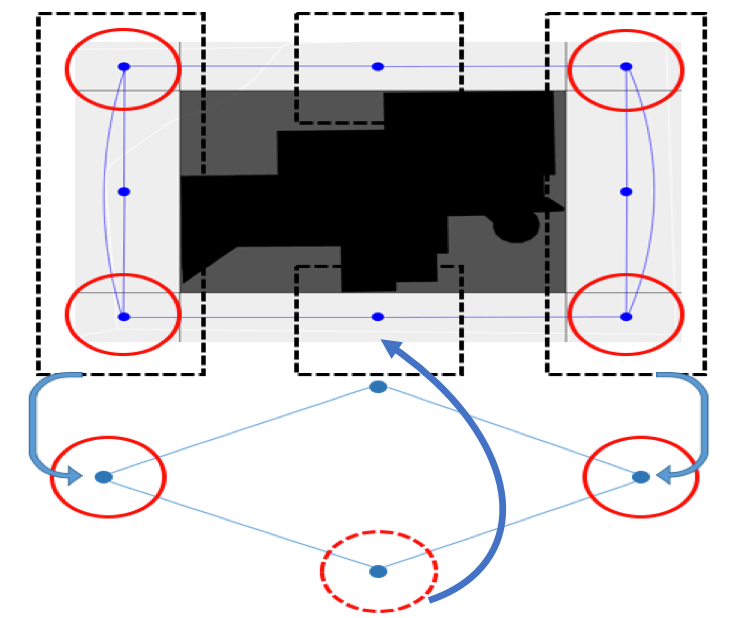}}

    \caption{(a)-(b) Visibility restrictions placed on the sensors in discrete space. We under-approximate visibility so the resulting assignment of discrete locations is valid for any real-valued location within the discrete region. (c) An example ``visibility graph", where each vertex represents a discrete region and edges exist between two vertices $u_1,u_2$ if the corresponding regions ``cover" each other, i.e. $\mathcal{C}'(u_1,u_2) = \textsc{true}$. (d) An example collapse of the connectivity subgraph $G_s$ into its connected components.  In this figure, the solid red circled vertices correspond to the vertices $\mathbf{S}$ selected by the MILP algorithm.  The connected components of $G_s$ dashed black boxes, and each is mapped to a collapsed node, with edges mapped to their corresponding collapsed nodes.  In this new graph, the Steiner tree problem solution identifies the vertex in a dashed red circle as the necessary node for connectivity.}
    \label{fig:visibilityrules}
\end{figure*}

\parae{Visibility Relaxation (C2).}
The visibility between a sensor and a location is given by \eqnref{eqn:visi}. However, to create a MILP encoding we must remove any non-linear constraints. To this end, we assume that a sensor, located in a space directly adjacent to an obstacle, cannot sense any discrete space at or past the obstacle's position.  For example, if a sensor is in a space diagonal to an obstacle, it cannot sense any space on the vertically or horizontally opposite side of the object (\figref{fig:visibilityrules:1} and \figref{fig:visibilityrules:2}). With this assumption we can pre-compute the visibility relation between any two grids in the system:
\begin{equation}
    \begin{split}
    \mathcal{B}'(u_i, u_j)  = \textsc{true}\  or \ \textsc{false}
    \end{split}
\end{equation}

%% The restrictions on sensor visibility due to obstacles are combined with natural sensor ranges $r_s$.  The sensed area is here under-approximated using the largest inscribed square within the circle defined by their sensing radius, and the use of a worst-case distance for coverage requires the sensing square to cover another location regardless of the sensor's placement within the discrete space $\mathbf{U}$.

% \pradipta{For further details on the method of computing these relations, please refer to the appendix.}

\parae{Modified Coverage Constraint (C1).} With the visibility relaxations, we can define a modified coverage function as:
\begin{equation}
\begin{split}
    \mathcal{C}'(u_i, u_j) &= \Big( \mathrm{dist}'(u_i, u_j) \leq r^s\Big) \bigwedge  \mathcal{B}'(u_i, u_j)  \\
\end{split}
\end{equation} 
where we use sensors of fixed sensing radius $r_s$ for ease of exposition (below, we describe how to generalize this to heterogeneous sensors).
To account for the discretization, instead of assuming that a sensor has visibility defined by the sensor range $r_s$, we assume its visibility is restricted to the largest inscribed square within the circle aligned to the grid axes (\figref{fig:visibilityrules}). This under-approximation ensures that we do not miss coverage, at the expense of optimality.

\parae{Transforming $k$-coverage into a graph vertex cover.}
With the modified coverage function, let us define a graph $G_v = (V, E_v)$ where vertex $i$ corresponds to a discrete region $u_i\in\mathbf{U}$.  We create an edge between vertices $i, j \in V$ when the corresponding discrete regions $u_i, u_j \in \mathbf{U}$ can ``cover'' each other, \ie $\mathcal{C}'(u_i,u_j) = \textsc{true}$.  \figref{vizgraph} illustrates this \textit{visibility graph}.  Now, \emph{the sensor $k$-coverage problem reduces to identifying if there exists a selection of vertices $S\subseteq V$ and their respective weights (number of sensors to place), $n_i \in \mathbb{Z}_{\geq 0}$, such that every vertex (i.e., every $u\in\mathbf{U}$) is covered by a set of neighboring vertices of $G_v$ with a total minimum weight of $k$ with a total budget of $\sum_{i\in V} n_i\leq N$.} 

% \begin{figure}[h]
% \centering
% \includegraphics[width=0.45\linewidth]{images/Obstacle_with_connectivity_edges.png}
% % \subfloat[]{\label{bipartite}\includegraphics[width=0.47\linewidth]{images/Bipartite.png}}
% \caption{An example ``visibility graph", where each vertex represents a discrete region and edges exist between two vertices $u_1,u_2$ if the corresponding regions ``cover" each other, i.e. $\mathcal{C}'(u_1,u_2) = \textsc{true}$.}
% \label{vizgraph}
% \end{figure}

A MILP formulation of $k$-vertex cover uses only $|V| = |\mathbf{U}|$ integer variables. Consider a integer variable $n_i \in \mathbb{Z}_{\geq 0}$ associated with each $i\in V$ which represents the number of sensors to place on that grid. Now, the $k$-cover goal becomes:
\begin{equation}
    \label{sumbools1}
     \sum_{j \in N_i(G_v)}  n_j  \geq k, \quad \forall i \in V
\end{equation}
where $\mathcal{N}_i(G_v) = \{ j \in V : (i,j) \in E_v \}$ denotes the neighbors of node $i \in V$ in the graph $G_v = (V,E_v)$. This implies that for every $u_i\in\mathbf{U}$, at least $k$ sensors is placed in its neighboring ($\mathcal{C}'(u_i,u_j) = \textsc{true}$) vertices. To ensure that we only place the maximum number of available sensors $N$, we need an additional constraint: $\sum_{i = 1}^{|V|} n_j \leq N$.
With this, the objective of the MILP formulation is:
\begin{equation}
\label{milp-form}
    minimize\ \sum_{i = 1}^{|V|} n_j 
\end{equation}

% \begin{equation}
% \begin{split}
% \label{milp-form}
%     minimize\ \sum_{i = 1}^{|V|} n_j \quad \mbox{Subject to:} &\sum_{j \in N_i(G_v)}  n_j  \geq k, \quad \forall i \in V \\
%     &\sum_{i = 1}^{|V|} n_j \leq N  \ \ ,\ \ \ n_i \in \mathbb{Z}_{\geq 0} 
% \end{split}
% \end{equation}

\parae{Accommodating Heterogeneity.}
If the network synthesis requires each location to be covered by multiple types of sensors such as cameras and microphones, we can treat each type of sensor placement problem as a separate coverage problem and solve using~\eqnref{milp-form}. If the synthesis specifies multiple sensors of the same type (e.g., camera) but with different specifications, we can generate a separate visibility graph for each sensor specification ($t \in \{1, \cdots T\}$) as $G^t_v = (V, E^t_v)$, introduce an integer variable $n_{t, i}$ to represent the number of type $t$ sensors placed in the discrete region $u_i$, and update \eqnref{sumbools1} and \eqnref{milp-form} in a straightforward way (details omitted) to formulate the synthesis.

%% \parae{Accommodating Heterogeneity.}
%% If the network synthesis requires each location to be covered by multiple types of sensors such as cameras and microphones, we can treat each type of sensor placement problem as a separate coverage problem and solve using~\eqnref{milp-form}.

%% % \emph{ Heterogeneity in Sensor Capabilities:} 
%% For network synthesis with multiple sensors of the same type (e.g., camera) but with different specifications, we generate a separate visibility graph for each sensor specification ($t \in \{1, \cdots T\}$) as $G^t_v = (V, E^t_v)$. 
%% Let us assume that we have a maximum of $N^t$ sensors of type $t\in \{1, \cdots, T\}$.
%% Then we can introduce integer variable, $n_{t, i}$ to represent the number of type $t$ sensors placed in the discrete region $u_i$. The $k$-cover and cardinality constraints then become: 
%% \begin{equation}
%%     \label{sumbools2}
%%     \begin{split}
%%         \sum_{t = 1}^{T} \sum_{j \in \mathcal{N}_i(G^t_v)}  n_{t,j}  \geq k, \quad \forall i \in V \ \ \  \mbox{and} \ \ \sum_{j = 1}^{|V|} n_{t,j} \leq N^t
%%     \end{split}
%% \end{equation}

% \begin{equation}
% \label{hetero-sat}
% \begin{split}
%     &\bigwedge_{i\in V}\quad\left( \ \sum_{t = 1}^{T} \sum_{j \in \mathcal{N}_i(G^t_v)}   \ \  \sum_{z = 1}^{Z^t} b_{t, j,z} \geq k \right) \ \ and  \ \sum_{j = 1}^{|V|} \sum_{z = 1}^{Z^t} b_{t,j,z} = N^t % \ \ \ \forall t \in \{1, \cdots, T\}
% \end{split}
% \end{equation}
%Better representation of such heterogeneity is left as a future work.

\parae{Obstacles and sensor placement (C4).}
The visibility graph accommodates the placement constraints by construction.

% \parae{$k$-Coverage with fewest sensors.}
% With the above-mentioned constraints, we can invoke modern SAT solvers such as z3~\cite{z3}.  The resulting SAT output will present either a set of feasible vertices (and therefore discrete locations $u\in\mathbf{U}$) for sensor placement or state that no such configuration exists.  Now to find the minimum number of sensor ($N$) to deploy, we do a binary search, starting from a very large number (typically the number of grids times k).

%% In each step, we check if a solution is feasible. If feasible, we reduce the the value of $N$ by half and check again. If not feasible, we choose a higher value in between the last known feasible value and the current value. This is repeated until we have a value of $N$ which results in a feasible solution and all values less than $N$ are infeasible. 

\parab{Determining connectivity (C3).} 
To capture the connectivity constraint, we can leverage a fundamental result that a graph $G$ is connected if and only if the second largest eigenvalue of the Laplacian matrix $L = D-A$, where $D$ is the degree matrix and $A$ is the adjacency matrix of $G$, is (strictly) positive. For a given graph (that is, with fixed nodes and vertices), checking whether this condition holds can be computationally challenging. In the topology synthesis problem further complications arise, as we are interested in imposing such a (nonconvex) connectivity constraint on graphs where the placement of nodes and edges is a design variable. Following the reduction proposed in~\cite{CDC_Bayen}, in a problem where the nodes and edges of a graph are given (and constant) the eigenvalue constraint can be expressed equivalently by a (convex) linear matrix inequality (LMI) in auxiliary scalar variables. Then, the connectivity of $G$ can be determined in polynomial time using any convex solver applicable to semidefinite programs~\cite{SDP_Vandenberghe_Boyd}. 

If the nodes are fixed but the edges of the graph are a decision variable (as in the case considered in \cite{CDC_Bayen}), then the problem becomes combinatorial, that is, a mixed-integer semidefinite program (MISDP), as the elements of the adjacency matrix $A$ are also decision variables of the problem. In our network synthesis problem, when the placement of sensors is also to be determined, the connectivity of the resulting network can be related to an LMI on a variable size matrix: its dimensions depend on the number of grid points with deployed sensors and the feasibility (or not) of edges depends on whether such edges exist in the connectivity graph.

Encoding such relationships in MILP is possible, at the expense of introducing a significant number of auxiliary integer variables. Alternatively, after some  algebra, one can obtain a MISDP, with constraints as described before and an objective that minimizes the number of sensors. But, lacking specialized MISDP solvers, we expect such a formulation to not scale well. So, we have resorted to
a further simplification: we relax the LMI constraint into affine inequalities\footnote{The LMI constraint is related to the positive semidefiniteness of a matrix. A diagonally dominant matrix, the space of which can be characterized by affine inequalities that can be part of a linear program, is positive semidefinite; the opposite does not necessarily hold.}.
This results in an MILP formulation for our problem, albeit one that may be conservative on highly connected graphs.

\parae{Scaling With Connectivity.} 
It turns out that using the Laplacian matrix for connectivity is not scalable beyond a very small problem space. Thus, we introduce another approximation method to simplify the connectivity as follows. We use a two-step  \textit{connectivity repair} algorithm to enforce connectivity after generating a topology satisfying the $k$-cover constraints. Similar to the creation of the visibility graph (see \figref{vizgraph}), we can create a \textit{connectivity graph} $G_c = (V,E_c)$, where vertex $i$ corresponds to a discrete region $u_i\in\mathbf{U}$, and edges exist between two vertices $i,j\in V$  if $\mathrm{dist}'(u_i, u_j) \leq r^c$. 

% Again, using this worst-case distance provides some robustness guarantees on any real placement of a sensor within the discrete region. One could also consider more complex connectivity models, where obstacles play a key role.
% \begin{figure}[t]
% \begin{center}
% \includegraphics[width=0.5\linewidth]{images/Collapse.png}
% \end{center}
% \caption{\label{collapse}An example collapse of the connectivity subgraph $G_s$ into its connected components.  In this figure, the solid red circled vertices correspond to the vertices $\mathbf{S}$ selected by the MILP algorithm.  The connected components of $G_s$ dashed black boxes, and each is mapped to a collapsed node, with edges mapped to their corresponding collapsed nodes.  In this new graph, the Steiner tree problem solution identifies the vertex in a dashed red circle as the necessary node for connectivity.}
% \end{figure}

\parae{A graph-based formulation for connectivity repair.}
A selection of sensors by the MILP coverage formulation, $\mathbf{S}$, induces a subgraph $G_s = (V_s,E_s) \subseteq G_c$ on the connectivity graph $G_c$.  $E_s = \{(i,j) \in E_c : \sum_{t = 1}^{T}  n_{t,i} \geq 1, \sum_{t = 1}^{T}  n_{t,j} \geq 1 \}$ \ie each grid has at least one sensor. To satisfy the connectivity constraint, $G_s$ must be a connected graph.  Should the MILP algorithm produce a connected $G_s$, no additional work is necessary.  Otherwise, we must add additional vertices to the set $\mathbf{S}$, playing the role of relay nodes.

To determine the relay node locations, we collapse $G_s$ into its connected components.  In this process, every single disconnected node is considered ``self-connected'' and mapped to an identical node in the collapsed graph (\figref{collapse}).  Edges from $G_c$ are similarly mapped to their corresponding collapsed counterparts.  The problem of identifying a set of nodes to add to $\mathbf{S}$ to make $G_s$ connected is identical to finding the minimal weight tree in the collapsed graph which spans the collapsed nodes of $G_s$.

This is the Steiner tree problem in graphs and is well-known to be NP-hard. We use a classical polytime approximation to solve the problem within a factor of $2(1-1/|\tilde{\mathbf{S}}|)$, where $|\tilde{\mathbf{S}}|$ is the number of collapsed nodes resulting from the initial selection $\mathbf{S}$~\cite{KouSteinerApproximation}. After adding the necessary nodes to create a tree in the collapsed graph to the initial set of nodes $\mathbf{S}$, the connectivity subgraph $G_s$ is connected.

% \parab{Modelling Sensor Heterogeneity.}
% With all the grid-based approximations, modeling heterogeneity becomes very straightforward and modular as follows. 

% \parae{ Heterogeneity in Sensor Types.} If the synthesis requires coverage by multiple types of sensors such as cameras and microphone, we can treat each sensor placement problem as a separate coverage problem. We can generate a separate visibility graph for each type ($t$) of the sensor, $G^t_v = (V, E^t_v)$ and solve the respecting k-vertex cover problem, followed by the connectivity repair method on the combined solution.

% \parae{ Heterogeneity in Sensor Capabilities.} 
% If the synthesis requires the placement of multiple sensors of the same type (e.g., camera) but with different specifications, we generate a separate visibility graph for each sensor specification ($t \in \{1, \cdots T\}$) as $G^t_v = (V, E^t_v)$. Then we can introduce a new Boolean $b_{t, j,z}$ to represent the placement of z-th type-t sensor in grid $j \in V$.
% \begin{equation}
% \label{hetero-sat}
% \begin{split}
%     &\bigwedge_{i\in V}\quad\left( \ \sum_{t = 1}^{T} \sum_{j \in \mathcal{N}_i(G^t_v)}   \ \  \sum_{z = 1}^{Z^t} b_{t, j,z} \geq k \right)\\
%     &\sum_{j = 1}^{|V|} \sum_{z = 1}^{Z^t} b_{t,j,z} = N^t  \ \ \ \forall t \in \{1, \cdots, T\}
% \end{split}
% \end{equation}
% where $N^t$ represents maximum number of sensor of type t, and $Z^t$ represent maximum number of type-t sensors allowed in same grid.  
% For connectivity, again, we can combine the solution and use the connectivity repair method.

\parab{The overall formulation.}
Summarizing our discussion so far, we can encode a MILP formulation of the k-coverage network synthesis goal with the following mapping of our four main constraints presented in \tabref{tab:prob_overall_sat}.

{\small
\begin{table}[htbp]
    \centering
    \caption{Summary of MILP Encoding of k-Coverage}
    % \resizebox{}{!}{
    {\begin{tabular}{ccc}
            \hline
        Coverage  & (C1)  &  k-Vertex Cover in Visibility Graph \\     
        Visibility & (C2)  &  Visibility Graph \\
        Connectivity & (C3)  &  Steiner Tree in Connectivity Graph\\ 
        Placement & (C4)  & Not Required\\ \hline
    \end{tabular}}
    \label{tab:prob_overall_sat}
  \end{table}
}

\section{Hierarchical Synthesis}
\label{s:hier-synth}

Both SMC and MILP, by themselves, fail to achieve our scaling goals (\secref{s:netw-synth-appr}). We apply hierarchical synthesis to achieve these.

\begin{figure*}[t]
    \centering
    \subfloat[]{\label{fig:hiera_ilustration}\includegraphics[width = 0.5\linewidth]{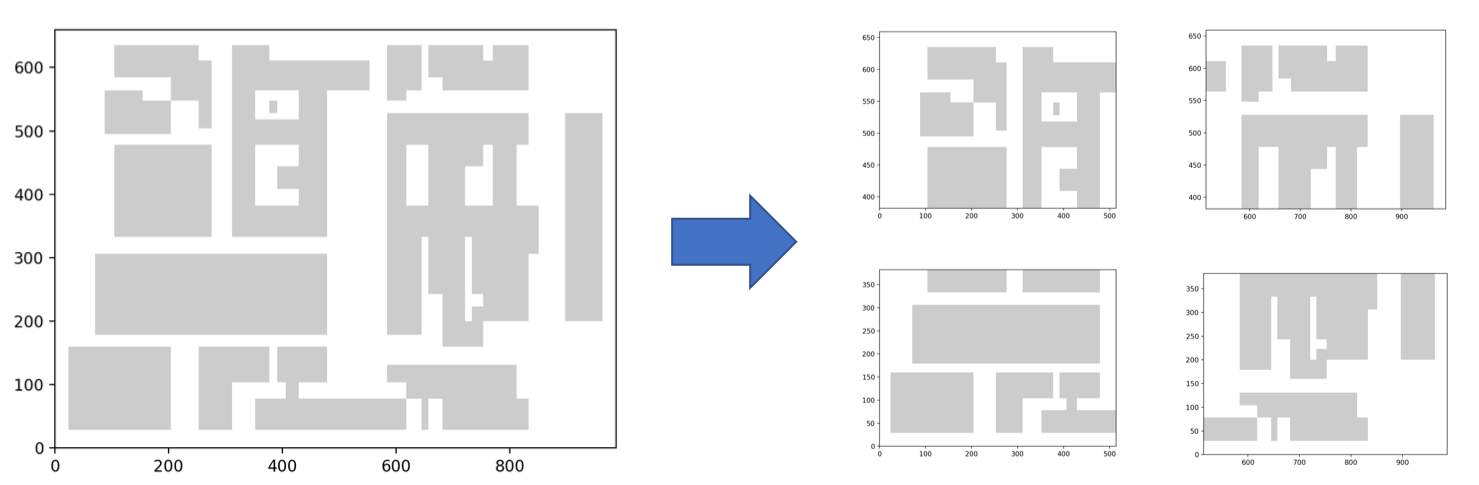}} \quad
    \subfloat[]{\label{fig:conn_reest}\includegraphics[width = 0.4\linewidth]{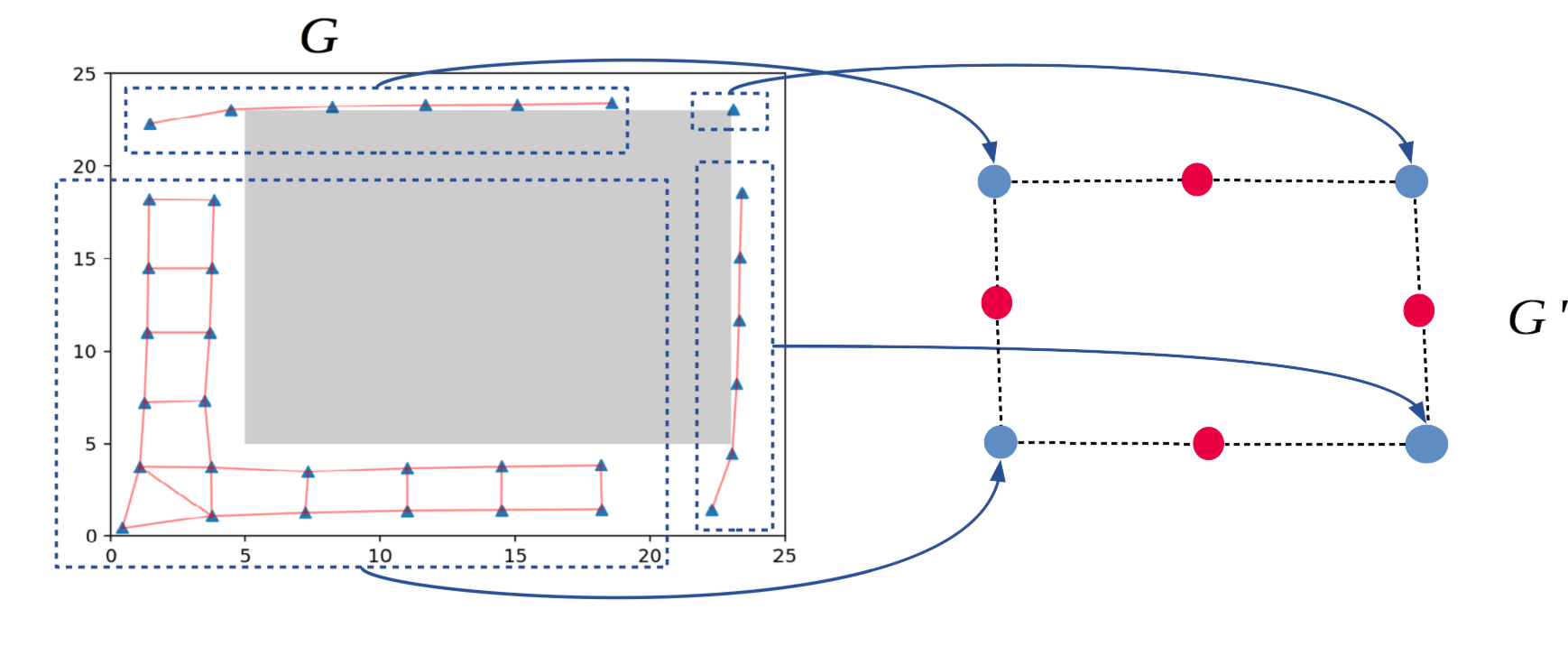}}
    \caption{ (a) Hierarchical synthesis solves sub-regions, then combines these solutions to satisfy coverage and connectivity over the entire deployment region using a (b) Connectivity repair method between sub-problems}
\end{figure*}

% \begin{figure}[t]
%     \centering
%     % \subfloat[]{\label{fig:conn_reest:1}\includegraphics[width = 0.5\linewidth]{images/subnetwork.png}}
%     \includegraphics[width = 0.9\linewidth]{images/collapsing.png}
%     \caption{Connectivity repair in hierarchical synthesis}
%     \label{fig:conn_reest}
% \end{figure}

%% One major issue with the above method is lack of proper encoding of physical distances between sensors. For example, in a symmetric network with sensing radius of $r_s$, if two sensed location is more than $2\cdot r_s$ distance apart, no sensor can sense both simultaneously. While the SMC method can figure these out via counter example generation, it takes much longer to identify all possible combinations. To circumvent, we precompute the pairwise distances between locations, identify the pairs with more than $2\cdot r_s$ distance, and add the necessary constraints. Similarly, in a symmetric network with communication radius of $r_c$, if two sensed location is more than $2\cdot r_s + r_c$ distance apart, two sensors that cover each of the location, can not communicate directly. Thus, there will not be any communication link between them. By adding such constraints, we are able to reduce the runtime and improve scalability. 

\parab{Hierarchical synthesis approach.}
In hierarchical synthesis, we subdivide the deployment area $\mathbf{L}$ into smaller sub-areas (\figref{fig:hiera_ilustration}) and solve for coverage (and not for connectivity) in each of the sub-areas separately. Then, once we have individual solutions, we need to connect them so that we have a connected network. For MILP, we use the graph based connectivity repair approach presented in \secref{s:basel-milp-form}. To connect the individual sub-problems in SMC, we first combine the solutions to check for connected pairs of nodes, then collapse each of the connected sub-networks into a single node (\figref{fig:conn_reest}). Then, we use constraints $\psi_4$ and $\psi_7$ in the SMC formulation to ensure connectivity between the sub-areas (similar to MILP connectivity repair but in the continuous domain). 

We could have solved for both coverage and connectivity in each of the sub-areas. In a complex deployment region, solving for both connectivity and coverage in each subproblem often over-provisions the network. Also, a $k$-coverage solution often results in connected sub-networks, so our approach enables faster synthesis by reducing computation (\secref{sec:experiment}).

\parab{Incremental Coverage Repair.}
Hierarchical synthesis can help scaling but may result in redundant coverage, because it solves each sub-region independently, so more than $k$ sensors may cover parts of the sub-region near the boundaries (\eg by sensors from neighboring sub-regions). To circumvent this, we first solve each sub-area for $k=1$ coverage. Then, for each grid element $u$, we measure the obtained coverage $k_u$ from this solution. The residual coverage requirement for $u$, then, is $k-k_u$; we now run the coverage problem again for each sub-area with these residual coverage requirements as constraints, which effectively ``fills'' up the coverage on grid points interior to the sub-area. After this, we apply the connectivity repair described above.

\section{Evaluations}
\label{sec:experiment}

In this section, we explore how the solution quality for MILP and SMC varies with problem size.

\subsection{Methodology}
\label{s:methodology}

%% \parab{Alternatives we compare.} We extensively compare our SMC and MILP formulations with the proposed hierarchy (for brevity, in the rest of this section, we refer to these simply as MILP and SMC).
%% For MILP we use hierarchy without the incremental coverage repair. The rationale behind that is for MILP the level of hierarchy is very shallow and thus over-provisioning via hierarchy is very small compared to SMC.

\parab{Implementation.} Our SMC implementation uses Z3~\cite{z3} for SAT solving, and CPLEX~\cite{CPLEX} for convex constraints. We use CPLEX also for the MILP formulation.

\parab{Inputs.} There are six inputs in our evaluation.
The \textit{deployment region} $\mathbf{L}$ denotes the discretized grid for which we synthesize the network. For most of our experiments, we use two different sizes of scenarios: (1) $20\times 20$ (\textit{Small}) and (2) $50\times50$ (\textit{Large}) grid. In the \textit{Small}, we use hierarchy with SMC but not with MILP since the latter scales to this scene. For the \textit{Large}, we use hierarchy for both since neither is able to scale to this problem size. The performance of synthesis algorithms depends heavily on the fraction of the deployment region occupied by obstacles (\textit{obstacle extent}), and on the spread of these obstacles across the deployment region (\textit{obstacle dispersion}). The \textit{coverage radius} $r_c$ and the \textit{communication radius} $r_s$ are also inputs; our results are sensitive to $\beta$, the ratio of the communication radius to the coverage radius. The \textit{grid granularity} controls how finely we can represent the deployment region; a finer grid implies lower coverage redundancy at the expense of scalability. Finally, we fix the coverage goal at $k=3$ (at least three sensors must cover each grid).

\parab{Scalability and Performance.} Since we are interested in synthesis for ad-hoc, time-constrained, deployments, we use the time taken to arrive at a good solution for a given problem size as a measure of scalability. Since we compare iterative synthesis methods, we specify this time as a constraint: we allocate a fixed amount of execution time $T_e$ on a fixed computing configuration for each of our methods and use the solution obtained by the end of that duration (or earlier if the synthesis has converged). A problem setting that does not scale will have arrived at a sub-optimal solution after $T_e$. In our experiments, we fix $T_e$ to 1 hour.

\parab{Metric.}
We then measure the performance of network synthesis by its \textit{coverage redundancy}. If the average number of sensors covering each location in the solution is $k_{avg}$, \textit{coverage redundancy} is $\frac{k_{avg}}{k}$ where $k$ is the desired coverage. For example, if, on average, 6 sensors cover every location, but we desire 3-coverage, then coverage redundancy is 2.

\begin{figure*}[!ht]
    \centering
    \subfloat[$\beta = 2, 20\times 20$ grid]{ \label{fig:heat1}\includegraphics[width=0.33\linewidth]{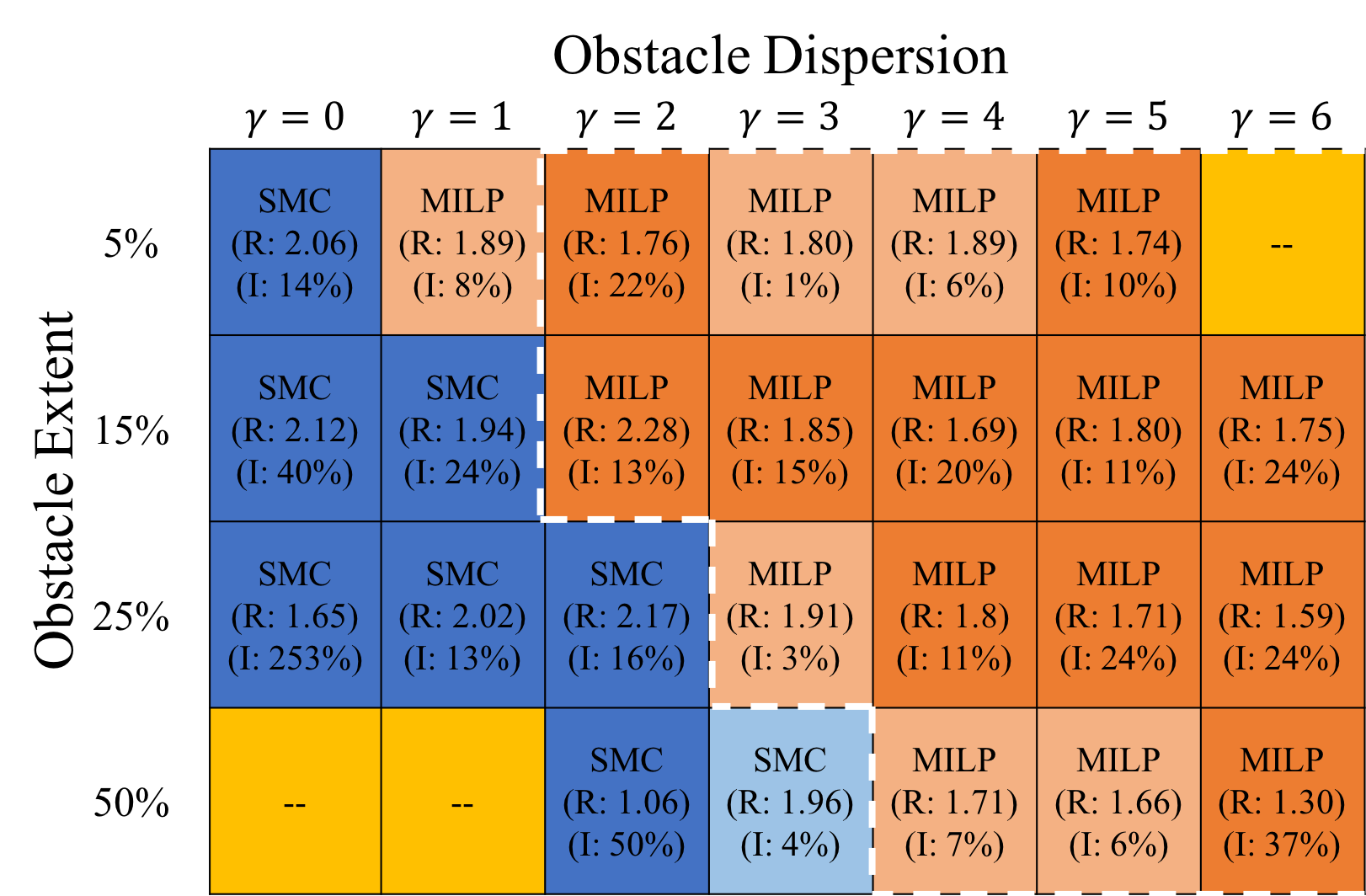}}
    \subfloat[$\beta = 1, 20\times 20$ grid]{\label{fig:heat2}\includegraphics[width=0.33\linewidth]{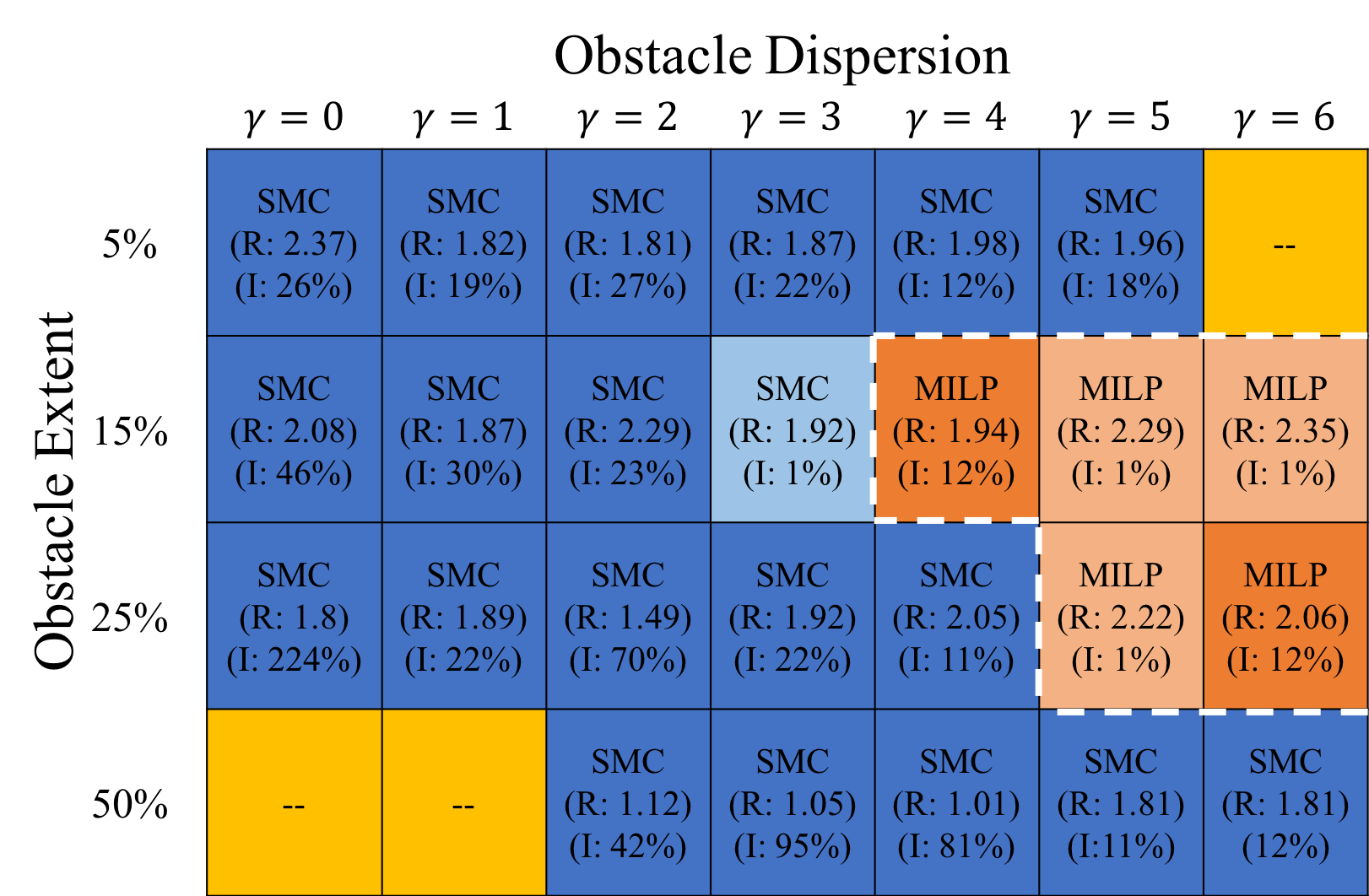}} 
    \subfloat[$\beta = 0.5, 20\times 20$ grid]{\label{fig:heat3}\includegraphics[width=0.33\linewidth]{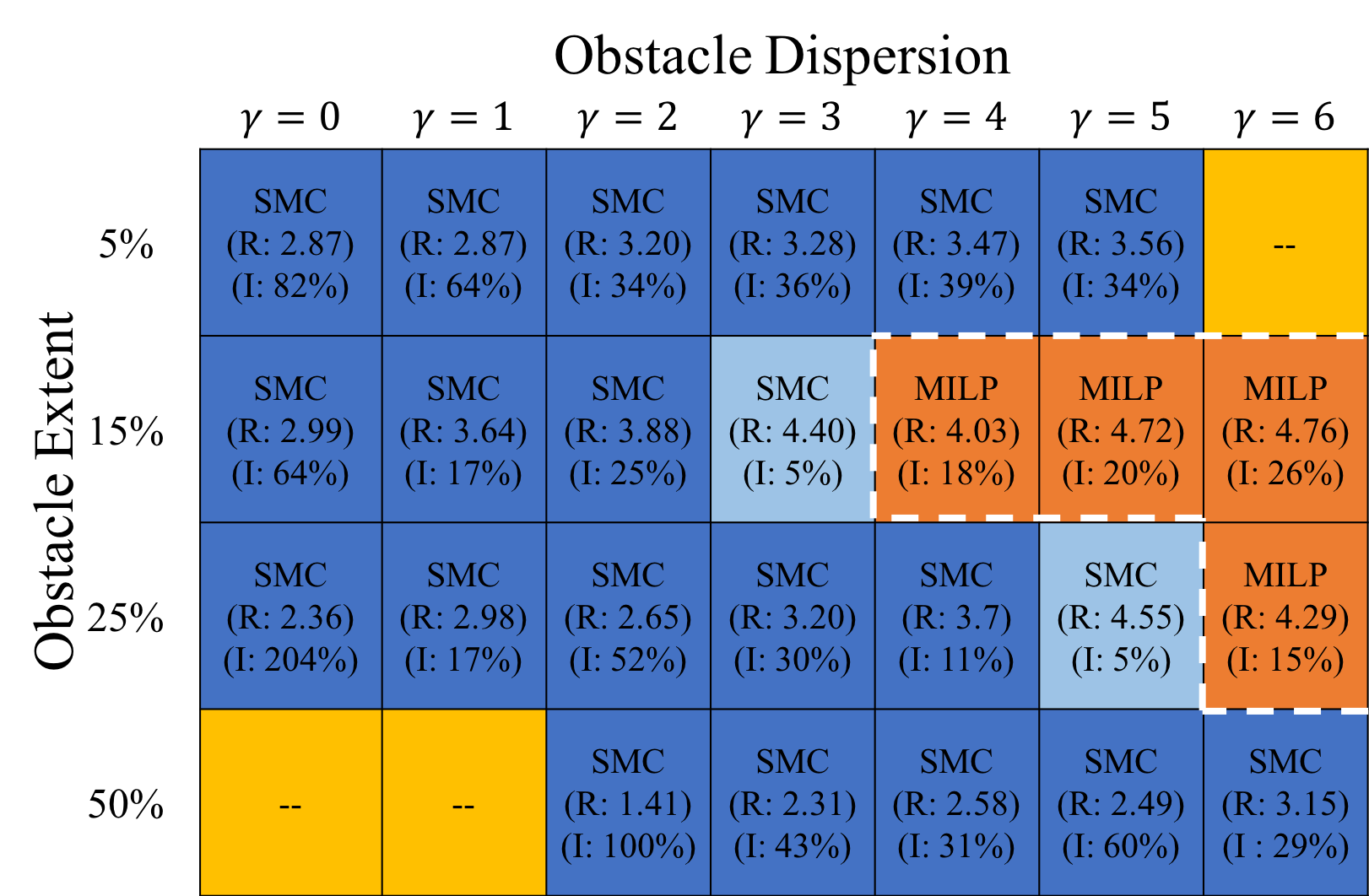}}\,
    \subfloat[$\beta = 2, 50\times 50$ grid]{ \label{fig:heat4}\includegraphics[width=0.33\linewidth]{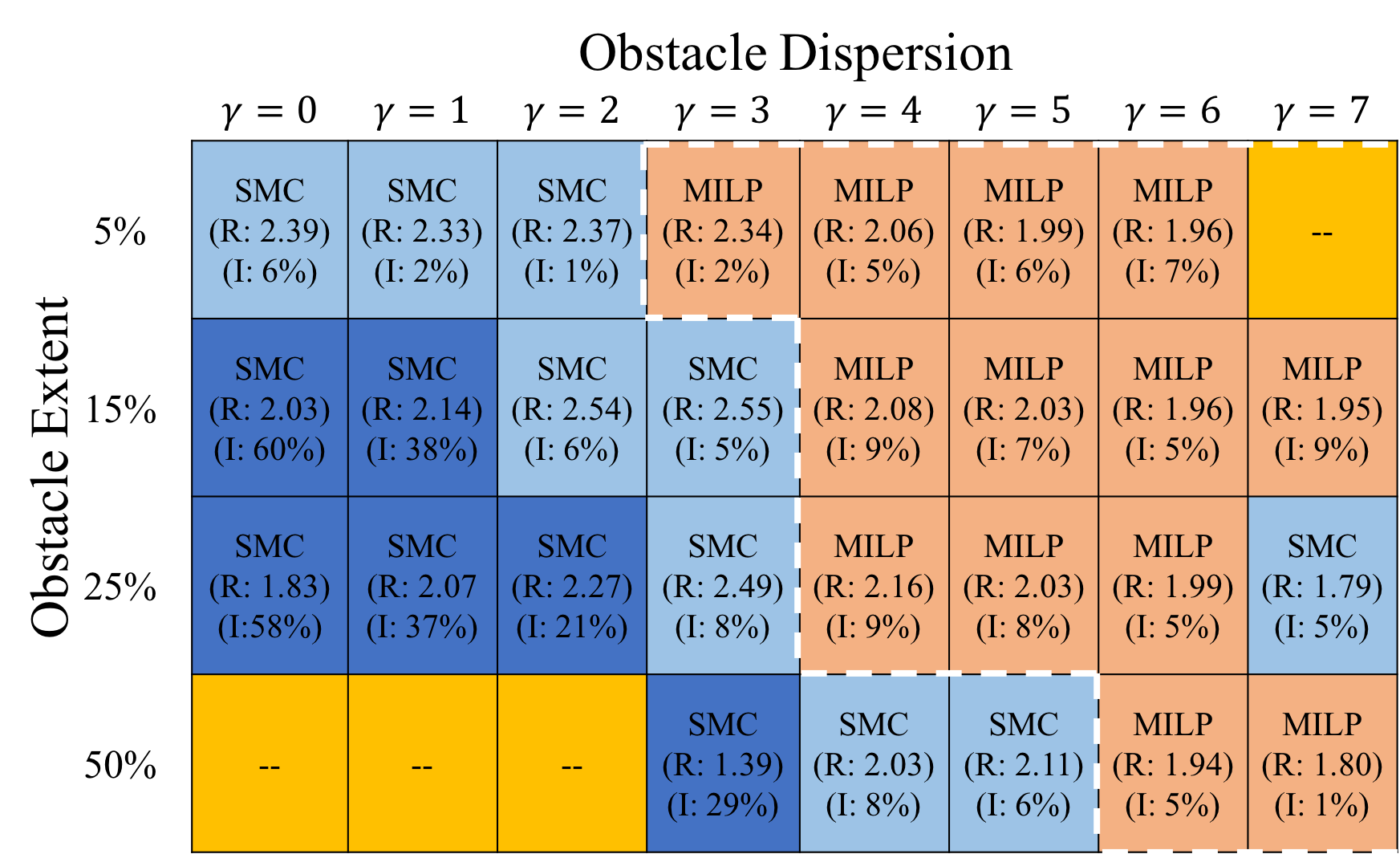}}
    \subfloat[$\beta = 1, 50\times 50$ grid]{\label{fig:heat5}\includegraphics[width=0.33\linewidth]{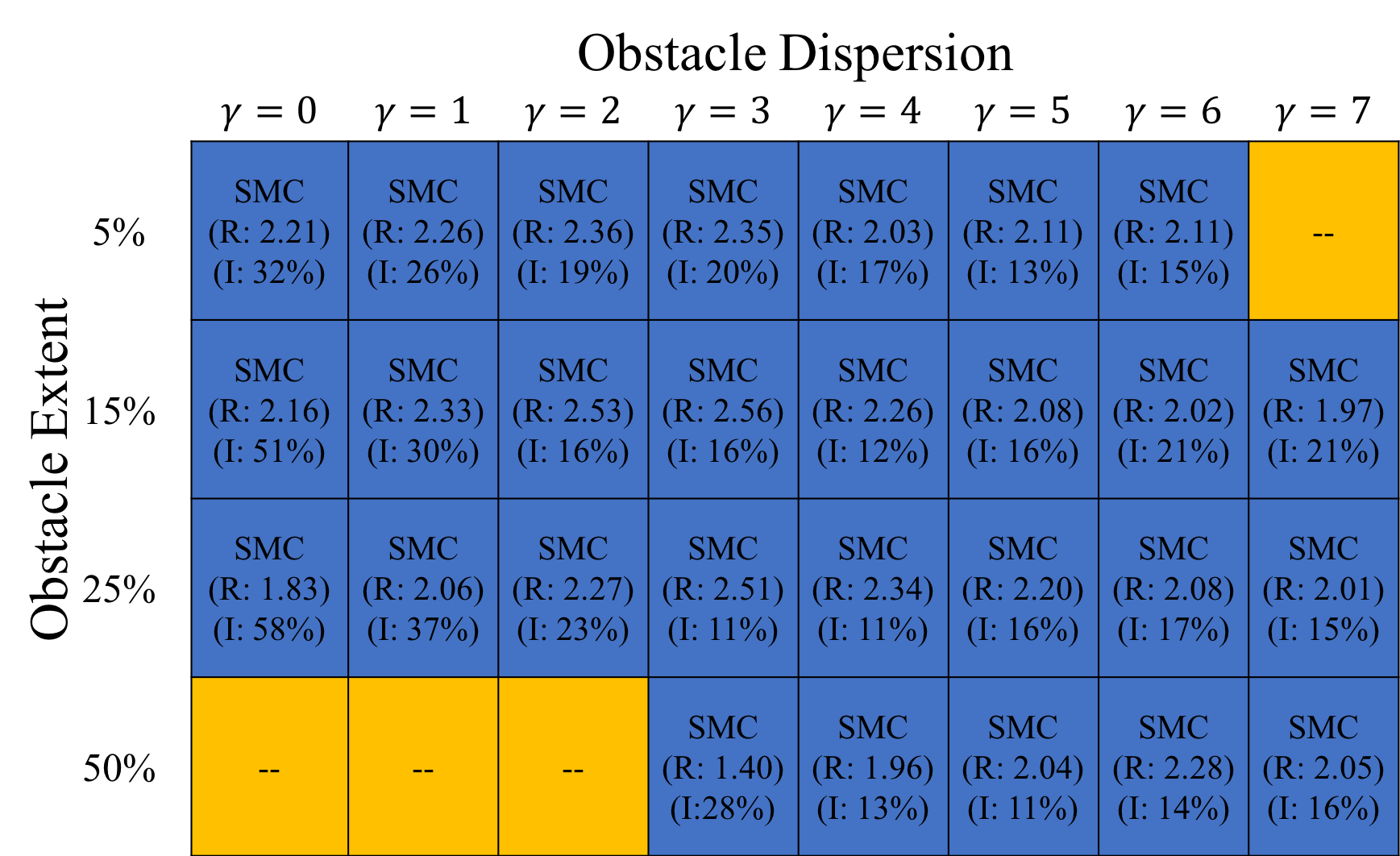}} 
    \subfloat[$\beta = 0.5, 50\times 50$ grid]{\label{fig:heat6}\includegraphics[width=0.33\linewidth]{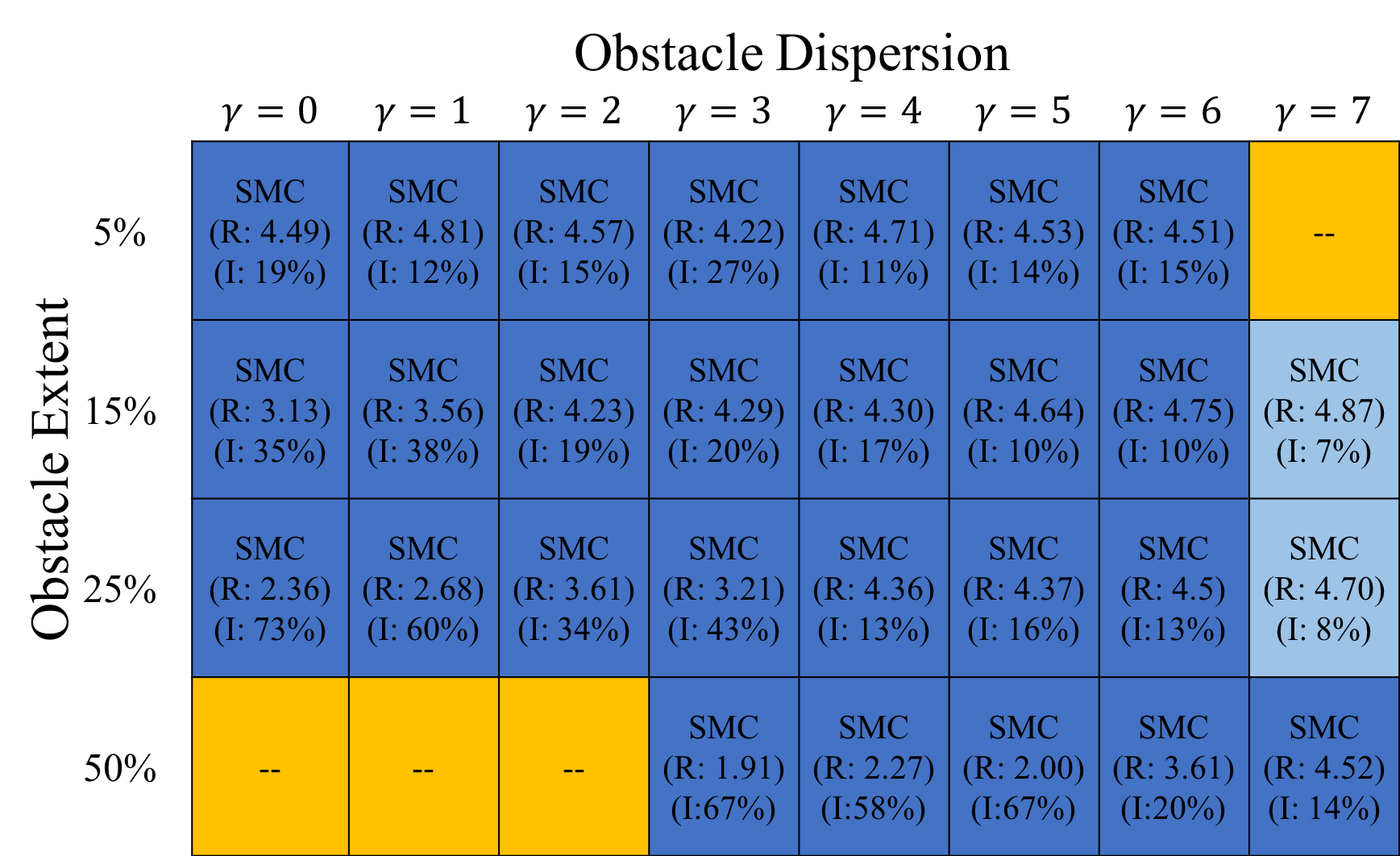}}
    \caption{Performance comparison for different combinations of obstacles extent, $\beta$ and $\gamma$. \textbf{Blue} $\implies$ SMC performs significantly better, \textbf{Light Blue} $\implies$ SMC performs slightly better ($<10\%$), \textbf{Light Orange} $\implies$ MILP performances slightly better (within 10\% of each other), \textbf{Orange} $\implies$ MILP performs significantly better, and \textbf{Yellow} $\implies$ infeasible test scenario.}
    \label{fig:heatmap}
\end{figure*}

%% The \textit{sub-area size} controls the size of subproblems in hierarchical approaches; smaller sub-areas scale better at the expense of coverage redundancy. \ramesh{We don't talk about sub-area size, should we remove it?}
%% \pradipta{Yes we can.}

%% This is only applicable to the SMC approach where smaller sub-area implies lesser complexity in sub-problems. However, this might lead to sensor over-provisioning. 
%% \end{enumerate}

% \ramesh{Change obstacle percent to obstacle extent}  

\subsection{The Role of Obstacles}
\label{s:role-obstacles}

Obstacles determine the visibility and placement constraints (\secref{sec:problem}) and significantly impact the performance of network synthesis. To evaluate how obstacle extent and obstacle dispersion affect the performance of MILP and SMC, we perform a set of experiments on synthetically generated obstacle distributions (in \secref{s:real-world-settings}, we explore realistic deployment regions).

To quantify the obstacle dispersion, we define $\gamma$ as the average number of adjacent grids with obstacles for each obstacle grid. $\gamma$ can theoretically vary from 0 to 8 with $\gamma = 7, 8$ being highly rare. For a fixed obstacle extent, \textit{lower values of $\gamma$ occur for small, widely dispersed obstacles}. We evaluate and compare the performance of MILP and SMC  for all feasible combinations of four different obstacle extents  5\%, 15\% 25\%, 50\%, and eight different values of $\gamma = \{0,1,2,3,4,5,6,7\}$ (We do not use $\gamma = 7$ for \textit{Small} as it is very rare).
In each case, we place obstacles on a $20\times20$ grid and a $50\times50$ grid and generate five obstacle placements (coverage redundancy varies little across placements, so we use 5 placements to minimize the total time to run experiments).
%% Random generation of multiple test-cases for each combination turned out to be non-trivial as certain values of $\gamma$ are prevalent for a fixed obstacle extent (e.g., for 5\% obstacle $\gamma = 0$ is most common). To circumvent this, we used a structured randomization of scenarios where we randomly place blocks of obstacles (with block size range as a function of $\gamma$).
We set the coverage radius to 6 units and explore the sensitivity of results to three different choices of $\beta = \{2, 1, 0.5\}$.  

\figref{fig:heatmap} presents the results of these experiments, both for \textit{Large} and \textit{Small}, where two methods are comparable if either their coverage redundancy 95\% confidence intervals overlap or the means are within 10\% of each other. These results indicate four distinct regimes of operation, discussed below.

\begin{figure}[!ht]
    \centering
    \subfloat[]{\label{fig:obspercentillus:2}\includegraphics[width=0.35\linewidth]{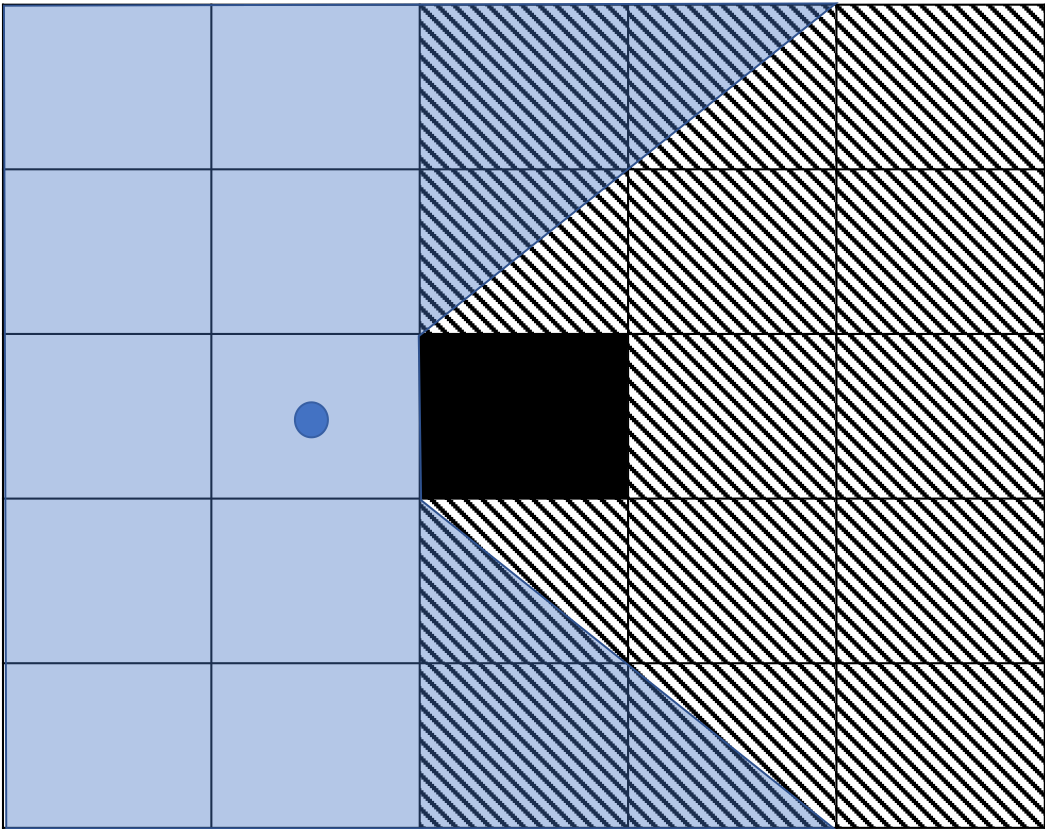}} \qquad
    \subfloat[]{\label{fig:obspercentillus:1}\includegraphics[width=0.35\linewidth]{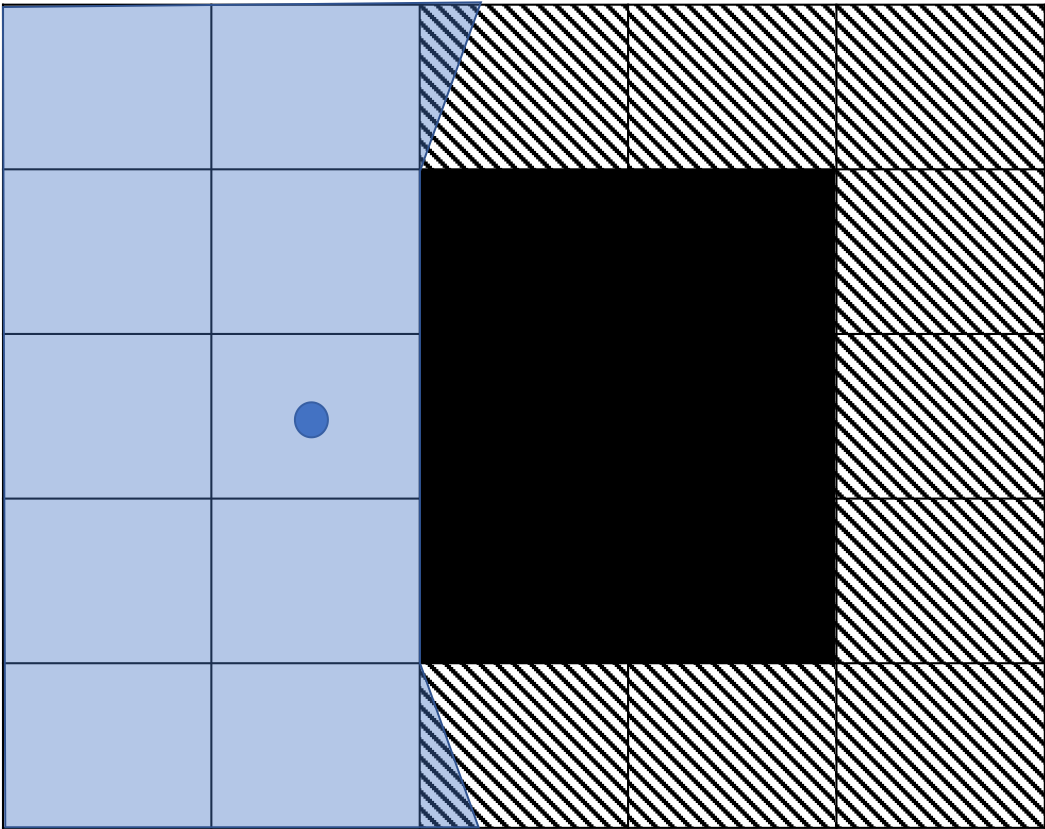}}

    \caption{Illustration of over-approximation for smaller $\gamma$ and larger $\gamma$. The hashed region illustrates the non-visible region in the MILP model and the blue shaded region is the original visible region.}
    \label{fig:obspercentillus}
\end{figure}

\parab{Small Obstacle Extent ($<15\%$).}
\figref{fig:heatmap} illustrates that for both \textit{Large} and \textit{Small}, with few obstacles (roughly $< 15\%$) SMC outperforms MILP for $\beta = 1, 0.5$. On the other hand, for $\beta = 2$ the performance largely depends on the obstacle dispersion (\figref{fig:heat1} and \ref{fig:heat4}) with better performance for MILP with lower obstacle dispersion ($\gamma > 3$). For $\beta = 2$, the coverage solution is also connected~\cite{xing2005integrated}; MILP converges faster to solutions with lower coverage redundancy in these settings both because its visibility approximation is less in obstructed environments (\figref{fig:obspercentillus:1}), and because the visibility graphs are less dense. It performs less well for $\beta = 1, 0.5$ because of its connectivity repair technique (\secref{s:basel-milp-form}).  For $\beta = 1$ (when the connectivity same as the sensing radius) and  $\beta = 0.5$ (when the connectivity radius is half the sensing radius), MILP has to deploy a large number of relay nodes to repair connectivity (made larger by the under-approximation of connectivity), which increases coverage redundancy significantly.

\parab{High Obstacle Dispersion ($\gamma \leq 3$).}
\figref{fig:heatmap} also illustrates that, regardless of the obstacle extent and value of $\beta$,  with dispersed obstacles (i.e., $\gamma \leq 3$) SMC outperforms or is comparable to MILP. With many small sized obstacles, MILP under-approximates visibility significantly (\figref{fig:obspercentillus:2}) and ends up placing more sensors. Thus, even with a large obstacle extent, if obstacles are small and widely dispersed, MILP's performance degrades.

\parab{Medium Obstacle Extent ($15\% -  25\%$) and Low Obstacle Dispersion ($\gamma > 3$).}
In this regime, MILP often outperforms SMC regardless of the values of $\beta$ for \textit{Small} when MILP does not require any hierarchy, but SMC does (\figref{fig:heat1}, \ref{fig:heat2}, \ref{fig:heat3}). In more obstructed environments, SMC requires a deeper hierarchy to scale well. In contrast, MILP converges faster to solutions with lower coverage redundancy in these settings both because its visibility approximation is less in obstructed environments (\figref{fig:obspercentillus:1}), and because the visibility graphs are less dense. However, as we introduce hierarchy to MILP (\textit{Large}), MILP's performance deteriorates due to over-provisioning as a result of hierarchy and SMC becomes the better alternative (\figref{fig:heat4}, \ref{fig:heat5}, \ref{fig:heat6}).

\parab{Large Obstacle Extent ($>25\%$) and Low Obstacle Dispersion ($\gamma > 3$).}
In this regime, the relative performance of these schemes again depends on the value of $\beta$ (similar to Small Obstacle Extent ($<15\%$)). For $\beta = 2 $, MILP is comparable to or better than SMC (\figref{fig:heat1} and \ref{fig:heat4}) and for $\beta = 1, 0.5$, SMC outperforms MILP (\figref{fig:heat2}, \ref{fig:heat3},  \ref{fig:heat5}, \ref{fig:heat6}). This results from the connectivity repair technique used in MILP. For $\beta = 2$, the coverage solution is guaranteed to be connected~\cite{xing2005integrated}.

\parab{Effect of $\beta$.} When $\beta=2$, a solution for coverage also results in a connected network. In contrast, when $\beta=0.5$, ensuring connectivity requires additional sensors. This is evident from the  coverage redundancy values in \figref{fig:heatmap}. For $\beta = 1, 2$, the best coverage redundancy is mostly within a factor of 2 of the optimal across the entire range of obstacle distributions while it is within a factor of 4 of the optimal for $\beta = 0.5$. In \figref{fig:heatmap}, we can see a nice separation between two regimes where SMC and MILP perform well for $\beta = 2$ (\figref{fig:heat1} and \ref{fig:heat4}). As $\beta$ goes from 2 to 1 to 0.5, the regime where MILP outperforms SMC vanishes. Since the primary difference between these regimes is the importance of connectivity, this shows that SMC handles connectivity constraints better than MILP.

\parab{Grid Dimensions.}
Finer grids can better model obstacles and reduce the under-approximation in visibility, but can reduce the scalability of synthesis. To understand this tradeoff, we explored the impact of changing the dimensions of each grid element, without increasing the number of grids. The generated heatmap (omitted for brevity) is consistent with the ones in \figref{fig:heatmap}. This suggests that the grid granularity affects both schemes equally, thus, has no effect on the boundary between SMC and MILP (\figref{fig:heatmap}).

\parab{Number of grid elements.}
As the number of grid elements increases (we fix the grid dimensions), MILP and SMC need hierarchy to scale. \figref{fig:heatmap} shows that for \textit{Small} there exists a difference between MILP and SMC (\eg \figref{fig:heat2}); MILP does not need hierarchy but SMC does. This difference disappears in the \textit{Large} (\figref{fig:heat5}). For MILP, the problem size beyond which it does not scale can be characterized by the number of unoccupied grids, $|\mathbf{O}|$. If this number is greater than a threshold $\chi$, then MILP does not scale without hierarchy because the number of constraints and solution search space increases dramatically (the number of constraints in MILP is a function of the number of open grids, since these determine its visibility and communication graphs, \secref{s:basel-milp-form}). $\chi$ is a function of the capacity of the solver~\cite{z3} and the computing configuration used for the synthesis. We can experimentally profile this quantity; in our experiments, $\chi$ is approximately 1200.

SMC requires hierarchy in both scenarios. Its performance improves relative to MILP, because MILP performance degrades because of the coverage and connectivity approximations. These arise from limitations in MILP's expressivity with respect to our network synthesis problem. This is evident from \figref{fig:heat5} and \ref{fig:heat6}: for the \textit{Large} with $\beta=1, 0.5$, SMC always outperform MILP significantly. For $\beta = 2$, we can see that the change is less severe as for $\beta = 2$ the coverage solution is the connectivity solutions and thus adding hierarchy deteriorates the performance of MILP slightly and makes it comparable to SMC.

To check whether the boundary between the two changes at larger settings, we performed a set of experiments with a $75\times75$ grid scenario and generated the respective heatmap (also omitted for brevity). This heatmap is comparable to the large results in \figref{fig:heatmap}. With increasing problem size, both methods appear to be equally affected by the addition of deeper levels of hierarchy needed to scale the synthesis.

% \begin{figure}[t]
%     \centering
%     \subfloat[5\% Obstacle Extent]{\label{fig:gridsize:1}\includegraphics[width=0.47\linewidth]{results/grid_size_errorstat_5.png}}
%     \subfloat[50\% Obstacle Extent]{\label{fig:gridsize:2}\includegraphics[width=0.47\linewidth]{results/grid_size_errorstat_50.png}}
%     \caption{Performance evaluation for varying grid granularity.}
%     \label{fig:gridsize}
% \end{figure}

\parab{Synthesis method selection.}
Taken together, these results suggest that MILP and SMC perform well in different regimes (indicated by the dashed line boundary in \figref{fig:heatmap}), so network synthesis should select which method to use depending on (a) the obstacle extent, (2) obstacle dispersion ($\gamma$), (3) the coverage to communication radius ratio ($\beta$), and (4) the number of open grids ($|\mathbf{O}|$). \tabref{tab:summary} summarizes these choices: use SMC except for $\beta > 1$ \textbf{and} $\gamma > 3$.

\begin{figure*}[t]
    \centering
    \subfloat[]{\label{fig:real:1}\includegraphics[width = 0.25\linewidth]{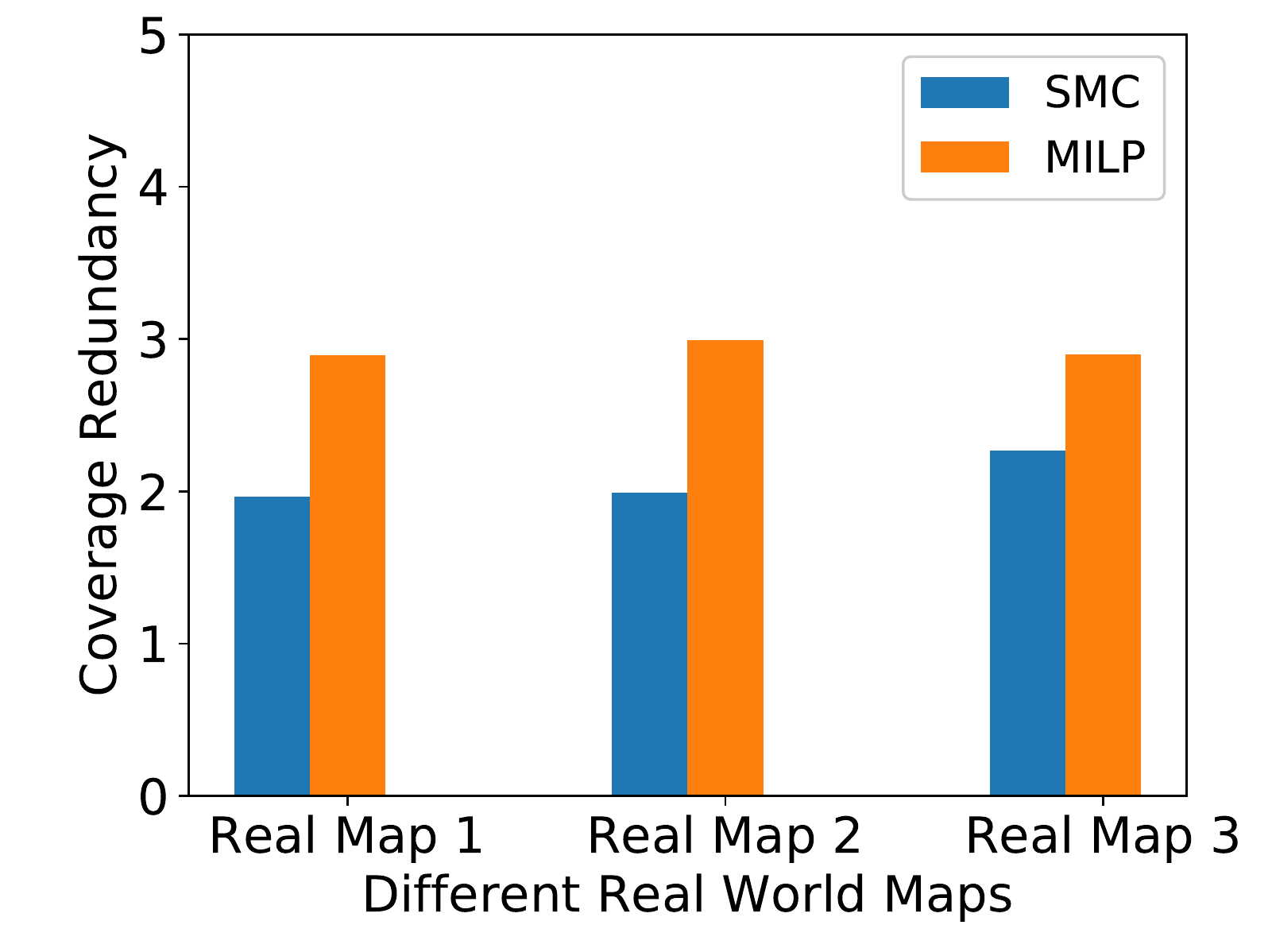}} 
    \subfloat[]{\label{fig:real:2} \includegraphics[width = 0.25\linewidth]{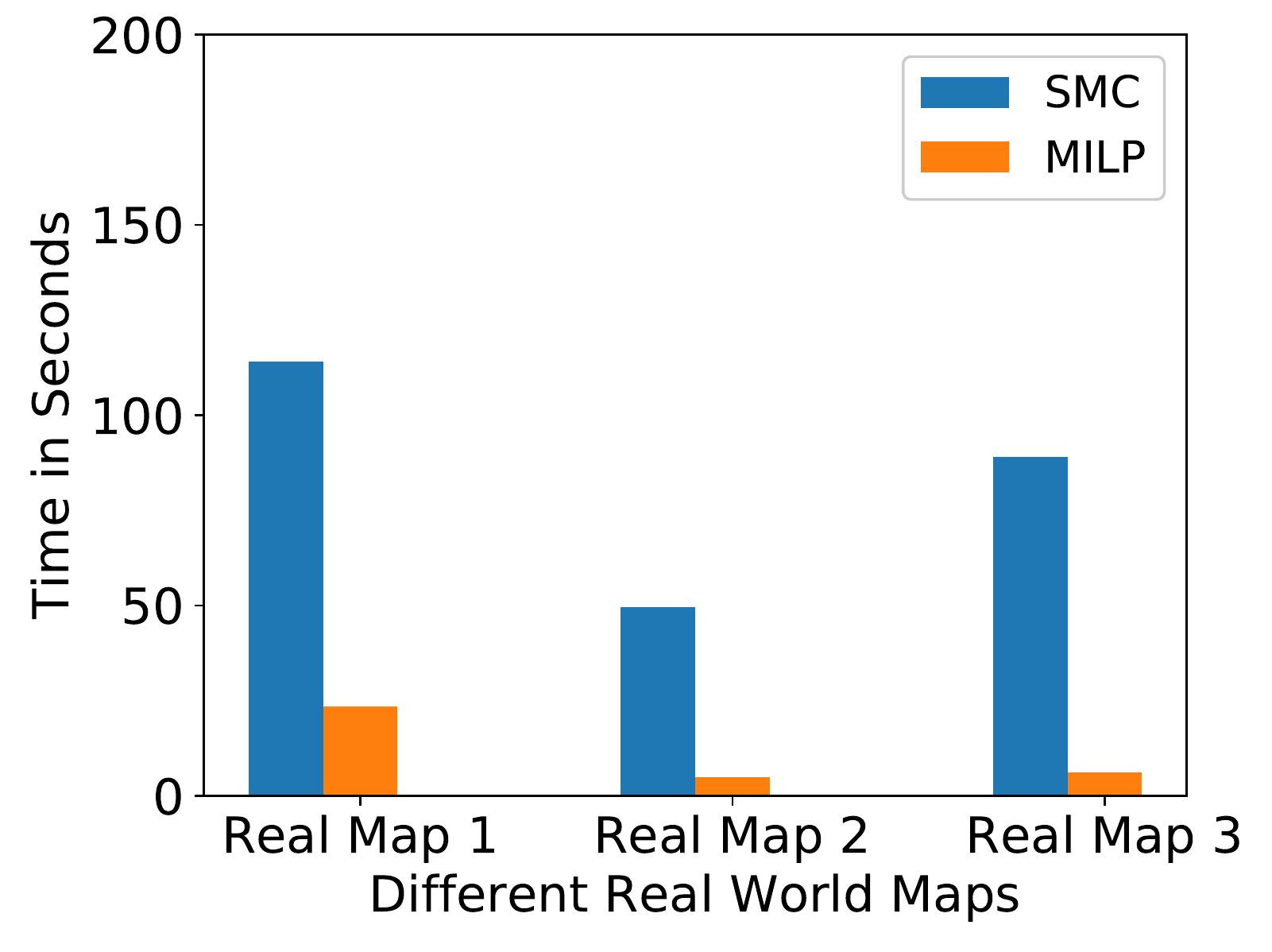}} 
    \subfloat[]{\label{fig:cover_hier}\includegraphics[width = 0.25\linewidth]{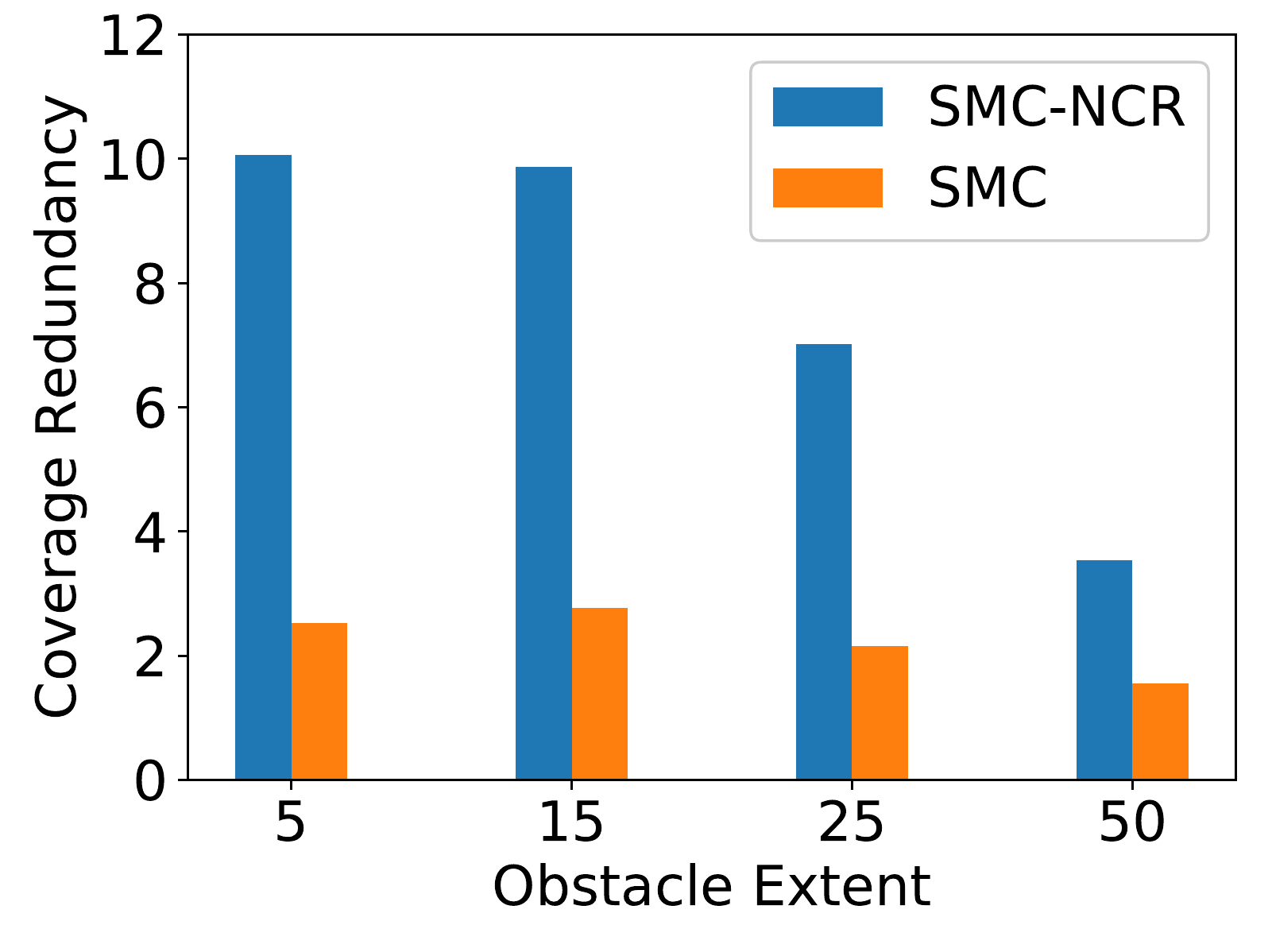}}
    \subfloat[]{\label{fig:smcconn:1}\includegraphics[width=0.25\linewidth]{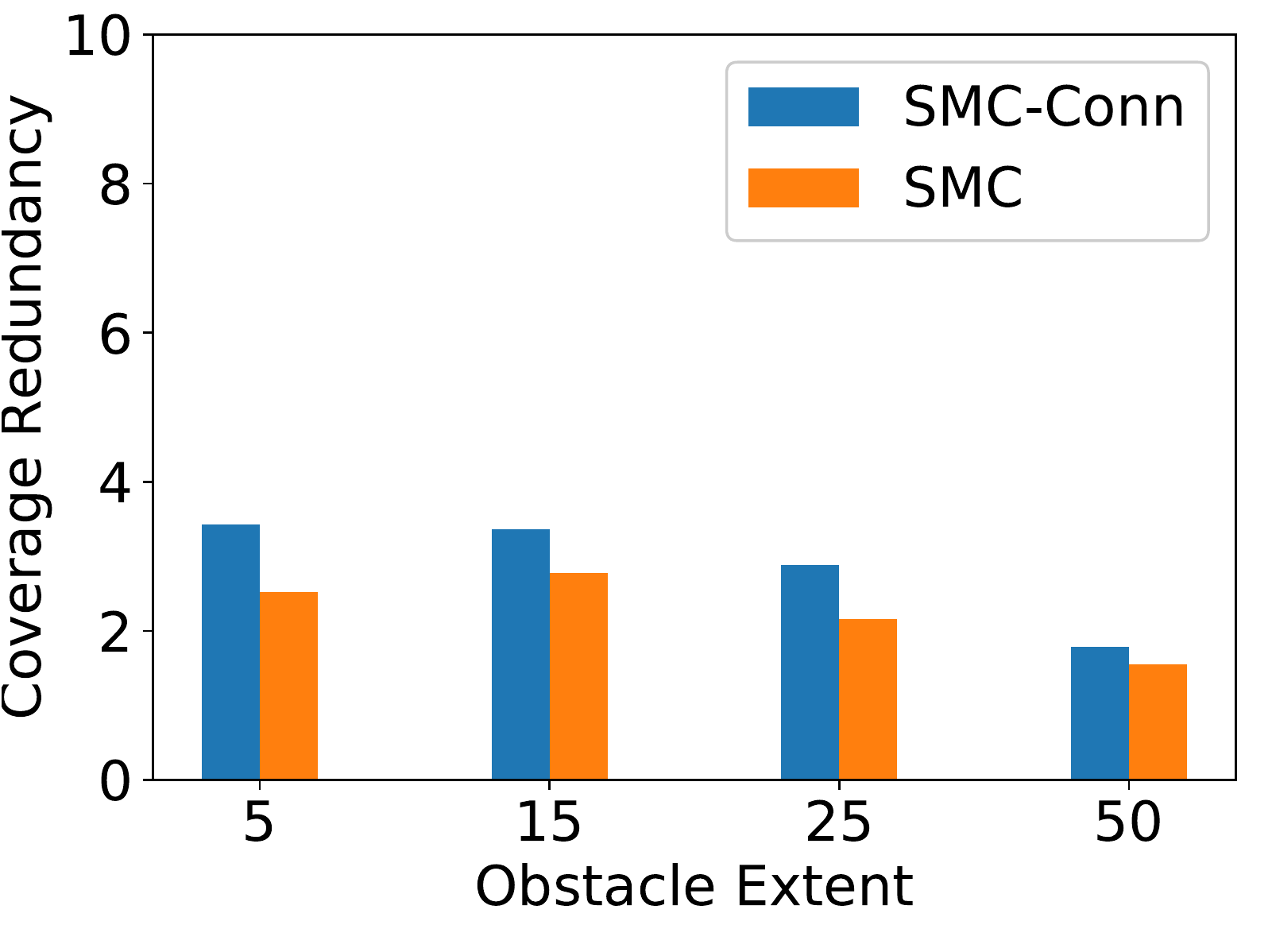}} 
    \caption{(a)-(b) Performance Analysis in real world dettings, (c) Comparison of performance for the proposed SMC with and without the incremental coverage repair, (d) Performance comparison between SMC and SMC-Conn for varying obstacle extents.}
    \label{fig:real}
\end{figure*}

\begin{table}[t]
    \centering
      \caption{Summary of the Experiments}

     \resizebox{0.9\linewidth}{!}{
    \begin{tabular}{|c|c|c|c|c|}
        \hline
        Obstacle  & Obstacle & $\beta$  &  \multicolumn{2}{c|}{Best Synthesis Method}  \\ \cline{4-5}
        Extent  &  Distribution $\gamma$ & &  $|\mathbf{O}| \leq \chi$ &  $|\mathbf{O}| > \chi$   \\ \hline
         Any &   $> 3$  & $ > 1$  & MILP  & MILP\\ 
         Any &   $\leq 3$  & $ > 1$  & SMC  & SMC \\
         $ < 15 \%$ &   Any & $\leq 1$  & SMC  & SMC \\
        $ \geq 15 \%$, $\leq 25 \%$   &   $> 3$ & $\leq 1$ & MILP/SMC  & SMC \\
        $ \geq 15 \%$, $\leq 25 \%$   &   $\leq 3$ & $\leq 1$ & SMC  & SMC \\ 
        $ > 25 \% $    &    Any & $\leq 1$ & SMC  & SMC\\ \hline
    \end{tabular}}
    \label{tab:summary}
\end{table}

\subsection{Real World Deployment Regions}
\label{s:real-world-settings}

To understand the relative performance of different approaches for network synthesis, we have presented results from synthetic deployment regions. In this section, we explore how synthesis scales to real-world deployment regions.

We extracted discretized representations from Google Maps for three large segments of university campuses in North America: (1) 740 m $\times$ 680 m with $37\%$ of obstacles with $\gamma = 5$ (2)  980 m $\times$ 640 m with $54\%$ of obstacles with $\gamma = 6$, (3) 860 m $\times$ 660 m with $35\%$ of obstacles with $\gamma = 5$. We use a grid size of $20~m\times20~m$ with a coverage and communication radius of 70~m following the standard ranges of Wi-Fi video surveillance cameras in outdoor settings. For each of these settings, SMC outperforms MILP (\figref{fig:real:1}) at the expense of higher runtimes (but still on the order of minutes, \figref{fig:real:2}) which is consistent with the findings in \figref{fig:heatmap}.

\subsection{Impact of Optimizations}
\label{s:impact-optimizations}

\parab{Synthesis with and without Incremental Coverage Repair.} Our hierarchical synthesis uses incremental coverage repair (\secref{s:hier-synth}): rather than solve for $k=3$ coverage in each sub-area, we first solve for $k=1$ since neighboring sub-areas can provide additional coverage. Then, we repair any missing coverage when combining the sub-areas.  \figref{fig:cover_hier} shows that incremental coverage repair can reduce redundancy by $2-4\times$ relative to a solution (SMC-NCR) that does not incorporate this technique.

%% we can achieve upto 2-3 times better performance compared to SMC without coverage hierarchy. This performance improvement is attributed to avoiding over-provisioning in each sub-problem. 
% \pradipta{Still running the experiment for 15\% obstacles.}

%% \parae{Takeaway.} Incremental Coverage Repair can improve the performance of original SMC upto 2-3 times  by avoiding over-provisioning.

% \begin{figure}[!ht]
%     \centering
%     \subfloat[]{\label{fig:cover_hier}\includegraphics[width = 0.5\linewidth]{results/obs_noopt_errorstat_1.eps}}
%     \subfloat[]{\label{fig:testbed} \includegraphics[width=0.47\linewidth]{images/IMG_4732.JPG}}
%     \caption{(a) Comparison of Performance for the Proposed SMC with and without the Incremental Coverage Repair, and (b) Real Testbed}
    
% \end{figure}

\parab{Solving for coverage alone.} Our synthesis method also incorporates an optimization where it does \textit{not} attempt to solve each subproblem both for connectivity and coverage, assuming that the coverage solution results in an almost always connected network. To understand the benefits of this approach, we compare SMC-based hierarchical synthesis against an alternative (SMC-Conn) in which we do solve for both connectivity and coverage in each sub-problem.
%%Our experiment settings are similar to those previous experiments:
%% RG: This will raise red-flags. I suggest you re-do the experiments after the deadline.
%% \added{(but with smaller scenarios of $20\times20$ grid space)}:
We explore four different obstacle extents, where each obstacle is the size of a single grid element and obstacles are uniformly distributed.

For these settings, \figref{fig:smcconn:1} shows that the optimization results in better quality solutions (lower coverage redundancy). This comes at the expense of slightly higher computation times: when the sub problem is not connected, SMC must repair its connectivity during the step in which it combines all subproblems, which increases the complexity of that step.

%% This set of experiments with five different percentage of obstacles (similar to \secref{sec:percent}) shows that SMC-Conn performs slightly worse than the proposed SMC method for all the test cases (illustrated in \figref{fig:smcconn:1}). This comparably inferior performance of SMC-Conn is contributed to over-provisioning of sensors with benefit of having  larger connected components prior to recovering connectivity. On the other hand, the proposed SMC method avoid over-provisioning in each subproblem and thereby requires slightly more time in connectivity reconstruction, illustrated in \figref{fig:smcconn:2}.

\parae{Takeaway.} Both our optimizations are important for reducing coverage redundancy and improving scalability.

% \begin{figure*}[t]
%     \centering
%     \subfloat[]{\label{fig:smcconn:1}\includegraphics[width=0.25\linewidth]{results/obs_percentage_errorstat_1_conn.eps}} \quad
%     \subfloat[]{\label{fig:smcconn:2}\includegraphics[width=0.25\linewidth]{results/obs_percentage_timestat_1_conn.eps}} \quad
%     \subfloat[]{\label{fig:testbed} \includegraphics[width=0.24\linewidth]{images/IMG_4732.JPG}}
    
%     \caption{(a)-(b) Performance comparison between SMC and SMC-Conn for varying obstacle extents, and (c)Real Testbed}
%     \label{fig:smcconn}
% \end{figure*}

\subsection{Incorporating Heterogeneity}
\label{s:incorp-heter}

\begin{figure}[t]
    \centering
    \subfloat[]{\label{fig:hetero} \includegraphics[width=0.5\linewidth]{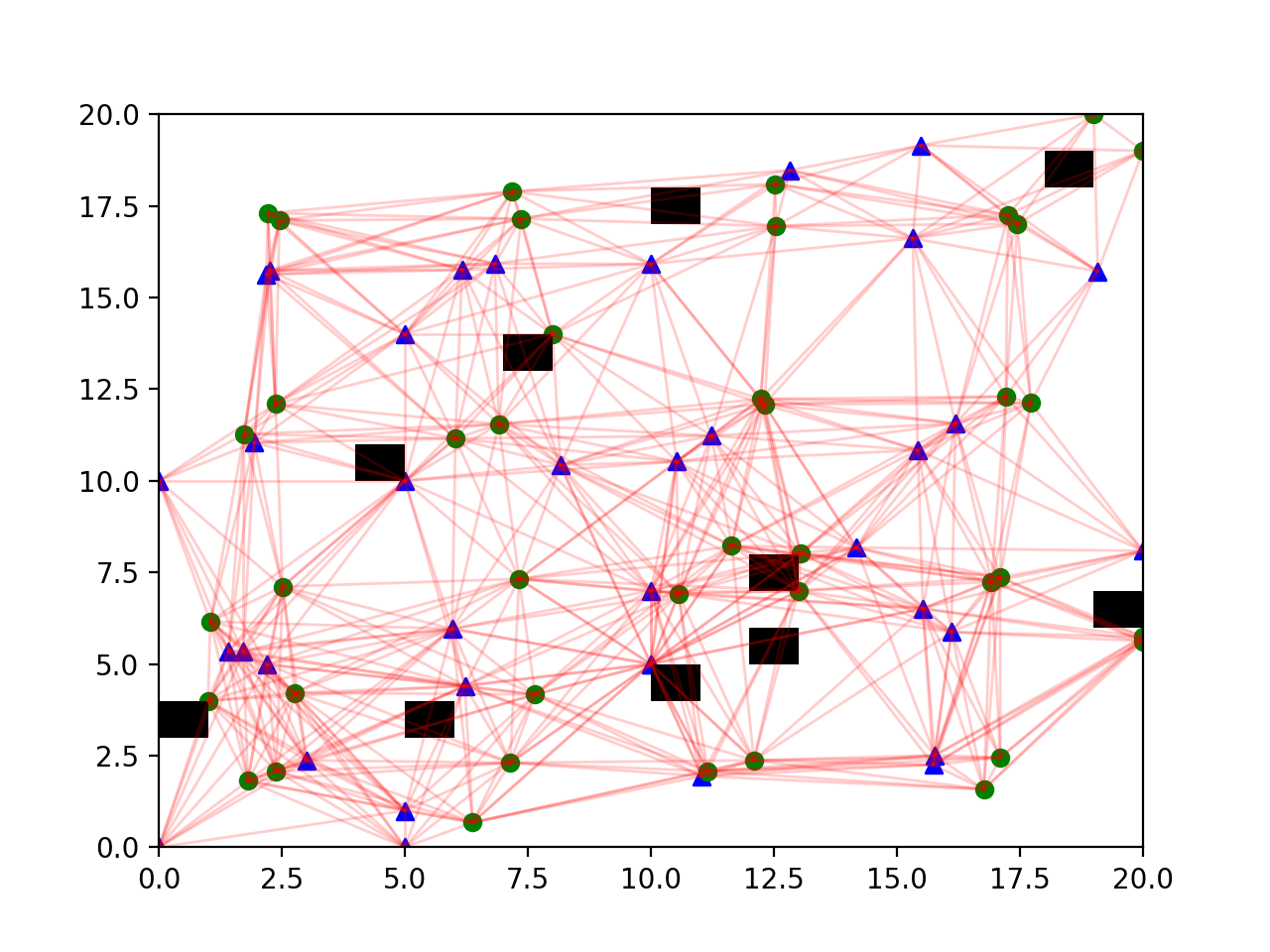}}
    \subfloat[]{\label{fig:testbed} \includegraphics[width=0.5\linewidth]{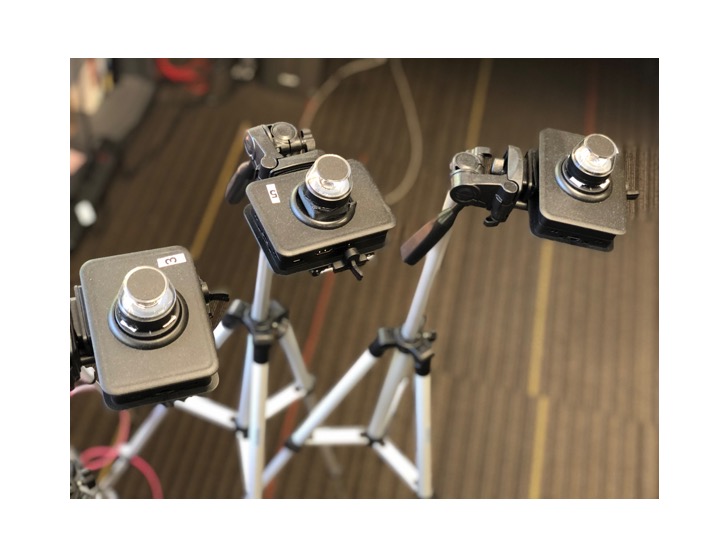}}

    \caption{(a) Heterogeneous sensor synthesis. The triangles represent one type of sensor, and the circles another. Each point is covered by one sensor of the first type and two of the second, (b) Real testbed}
    
\end{figure}

SMC and MILP can be easily extended to incorporate additional constraints. To illustrate this, suppose we want to synthesize a network with two different sensors (\eg camera and microphone), each with different coverage radii, with the additional constraint that every point in the deployment region should be covered by at least one sensor of the first type, and two of the second type. If both sensors are omnidirectional (we have left to future work an exploration of directional sensors), both SMC and MILP can be trivially extended to accommodate such constraints. \figref{fig:hetero} shows an SMC-synthesized network that satisfies such constraints.

\subsection{Testbed Validation}
\label{s:small-scale-valid}

The goal of this paper is to understand the scalability of network synthesis. A key question in network synthesis is \textit{validation}: when deployed, does the synthesized network actually satisfy the coverage and connectivity constraints? This is hard to do at large scales (\eg a campus) since it would require hundreds of sensors.

For this reason, we present a small-scale validation of network synthesis. We ran the synthesizer on a small deployment region, deployed a network corresponding to the output of the synthesis, and empirically determined if the connectivity and coverage requirements were satisfied. To do this, we used a testbed of 5 Raspberry Pi 3's with omnidirectional cameras mounted on top (\figref{fig:testbed}). The camera has a range (the distance at which a human can recognize a small object of standard dimensions, such as a cup) of $\approx 7\ m$. The Wi-Fi range of the device is roughly 15 meters.

We performed two experiments, one indoor, and one outdoor. In each case, we fed a discretized representation of the deployment region to the synthesizer, then placed the Raspberry Pis as directed by the synthesis output. To place the nodes accurately, we relied on relative measurements from the walls in the indoor settings and on GPS in the outdoor setting.

In the indoor experiment, the synthesized network \emph{Satisfied the  coverage and connectivity requirements.}

Outdoor validation was trickier for two reasons. First, the outdoor deployment region was heavily occluded, and GPS errors hampered our ability to place the sensors at exactly the locations specified by the synthesizer (GPS is known to be inaccurate in urban canyons~\cite{Liu18b}). The MILP formulation permits flexibility in placement within an entire grid; a MILP-synthesized network satisfied coverage. Second, the effective wireless range outdoors was smaller than that indoors, so we had to carefully profile the range and synthesize a network corresponding to this range. With this change, we were able to validate connectivity outdoors as well.

In general, with conservative choices of sensing and communication range, we expect a synthesized network to satisfy coverage and connectivity requirements when deployed. Any synthesis technique will need associated tools that can determine practical and safe values for sensing and connectivity radii, and we have left such tools to future work.

\section{Related Work}
\label{sec:related}

In this section, we briefly detail how the proposed network synthesis methods relate to the existing literature.

%% The three main classes of relevant literature are: \emph{Wireless Sensor Network, Camera Deployment, and Wired Network Synthesis.}

\parab{Wireless Sensor Networks.}
Over the last two decades, researchers have studied sensor network placement problems~\cite{younis2008strategies,wang2007efficient}. 
Common approaches rely on random dense deployment followed by a careful selection among the deployed sensors to fulfill the sensing goal~\cite{bai2006deploying}: often, this work has not considered obstacles. We highlight the most relevant prior work in this area: \cite{wang2011coverage,zhu2012survey,al2017optimal} present a more complete treatment.

\parae{Discrete Set of Sensors and Locations.}
With a set of randomly pre-deployed sensors, the network topology design reduces to controlling the sleep pattern of the sensors~\cite{santi2005topology,xing2005integrated,zhang2005maintaining, MCLCT, DisjointKCover, LifetimePlacement}. This can be easily encoded using SAT~\cite{SATdisjointkcovers} or framed using computational geometry~\cite{configurations}. None of these approaches deals with obstacles and visibility constraints for sensing.
% However, it narrows the focus of the problem in focus is much narrow and does not deal with area coverage or environment modeling. 
In contrast, we focus on a methodical top-down synthesis and support richer sensing goals such as $k$-coverage. Moreover, unlike these approaches, we focus on scaling to large deployments in highly occluded settings.
% % The work presented in \cite{LifetimePlacement} considers the problem of minimizing the number of sensors required to cover a given set of target locations while maintaining a network lifetime of a certain guarantee.  However, they only considers connectivity to a set of base stations and uses computational geometry tools for optimal placement in an obstacle-free environment.
% In \cite{configurations}, the authors analyzed configurations and optimal lattice structures in free space to achieve coverage and connectivity for sensors in uncluttered settings.
 
\parae{Submodular Optimization.} Prior work has explored a greedy algorithm for submodular optimization of sensor placement and scheduling~\cite{TrafficSchedPlace}. The proposed method does not consider connectivity constraints. Similarly, \cite{WaterSensors} uses greedy heuristics to solve several problems in selecting an optimal placement of sensors in a water network with submodular objectives. While sub-modularity assumptions enable rapid synthesis, they lack the expressivity of SAT and SMC which are necessary to jointly address coverage and connectivity constraints.
%% In contrast, the proposed methods can easily incorporate a wide range of sensing and communication requirements via representing them as additional Boolean or pseudo-Boolean constraints. 

\parae{MILP.} Prior work has surveyed and combined MILP/ILP approaches to solve the sensor placement problem~\cite{UnifiedApproach}. Our MILP formulation extends this with connectivity constraints.

%% This is also verified in our implementation of MILP based network synthesis.

\parae{Relay Placement.}
Prior work has also explored connectivity repair using relays~\cite{RelayNodes,Santos:2014}. In~\cite{FaultTolerant}, the authors address the problem of placing relay nodes to create k = 1, 2, and higher connectivity in both one and two-way communication scenarios. The problem is  NP-hard, and a series of polynomial approximation algorithms are presented to solve the problem for k = 1, 2 and a generalization is given for k > 2. In contrast, we do not focus on just relay placement, but on the synthesis of a connected network satisfying coverage objectives, a much harder problem. In~\cite{Santos:2014}, a set of relays is placed to establish a communication backbone. They consider obstacles and compute an Euclidean Obstacle-Avoidance Steiner tree to place the relays but this solution does not scale. 
    
\parab{Camera Placement.} There exists a body of work on camera placement  related to our work. Solutions to the well-known art-gallery problem for placing visual sensor~\cite{gonzalez2001randomized} generally assume infinite sensing range, and do not consider connectivity. One work~\cite{huang2014connected} explores an indoor camera placement problem with connectivity constraints, but in an uncluttered environment and assumes unlimited sensing range. Another work~\cite{erdem2006automated} considers limited camera ranges and polygonal obstacles, but does not consider connectivity. Other prior work has used quadratic convex programming~\cite{ghanem2015designing}, integer linear programming~\cite{yabuta2008optimum}, and linear programming~\cite{horster2006optimal} for camera placements, but again do not consider connectivity.

\parab{Wired Network Synthesis.}
Finally, prior work has explored synthesis of network topologies and routing configurations~\cite{beckett2017network,schlinker2015condor}. Our problem is qualitatively different in that it includes coverage requirements and visibility constraints.

\section{Conclusion}
\label{sec:concl}

Motivated by advances in solver technology, this paper considers the top-down synthesis of large-scale ad-hoc IoT networks. We explore two qualitatively different formulations: one represents synthesis constraints in the SMC framework that permits convex constraints, and another uses a conventional MILP optimization formulation. We also develop a hierarchical synthesis technique to scale to large problem sizes. Our results show that, somewhat surprisingly, SMC's solution quality is better than MILP at larger problem sizes, despite being fundamentally less scalable than MILP. This is likely due to the fact that MILP can only approximately capture coverage and connectivity constraints, and these approximations result in higher coverage redundancy. Several directions of future work remain including: extending the synthesis to accommodate directional sensors, and determining scaling techniques for these; incorporating computational elements into the network synthesis to place fusion nodes that can optimally process sensor data; and devising more scalable MILP formulations.

%% MILP does not scale beyond very small problems, SAT and SMC, when combined with 
%% between the regime of deployment specifications (in terms of obstacle extent and dispersions) where SAT or SMC performs considerably better. This separation is dictated by the communication to coverage radius ratio, $\beta$. With this, we developed a set of guidelines to choose the proper tool (SMC/SAT) for each sub-problem to develop a mixed hierarchical approach; this achieves low coverage redundancy with moderate runtime. We also presented a discussion on incorporating heterogeneity in the network synthesis.

% \begin{enumerate}
%     % \item Grid based Coverage and Spanning Tree Based Connectivity
%     \item Grid based Coverage and SMC Based Connectivity
%     \item Grid based Coverage and Connectivity
%     \item SMC based Coverage and Connectivity in Grid Space
%     \item SMC based Coverage and Connectivity in Continuous Space
%     \item Hierarchical Approach for all of the above
% \end{enumerate}

%\bibliographystyle{unsrt}  % INFOCOM papers appear to follow the IEEETran bibliography template (i.e. only initials from first name are displayed)
\bibliographystyle{ACM-Reference-Format} % this looks more like recently published INFOCOM papers
\bibliography{ref}
\end{document}